# Engineering of robust topological quantum phases in graphene nanoribbons

Oliver Gröning[1*], Shiyong Wang[1*], Xuelin Yao[2*], Carlo Pignedoli[1], Gabriela Borin Barin[1], Colin Daniels[3], Andrew Cupo[3], Vincent Meunier[3], Xinliang Feng[4], Akimitsu Narita[2], Klaus Müllen[2], Pascal Ruffieux[1] & Roman Fasel[1,5]

[1]Empa, Swiss Federal Laboratories for Materials Science and Technology, 8600 Dübendorf, Switzerland

[2]Max Planck Institute for Polymer Research, 55128 Mainz, Germany

[3]Department of Physics, Applied Physics, and Astronomy, Rensselaer Polytechnic Institute, Troy, NY 12180, USA

[4]Center for Advancing Electronics Dresden & Department of Chemistry, Technische Universität Dresden, 01062 Dresden, Germany

[5]Department of Chemistry and Biochemistry, University of Bern, 3012 Bern, Switzerland

[*]These authors contributed equally to this work.

**Topological phases of matter support robust, yet exotic quantum states at their boundaries such as spin-momentum locked transport channels or Majorana fermions [1-4]. The perspective to use such states in spintronic devices or as Qubits in quantum information technologies is a strong driver of current research in condensed matter physics [4-6].** The understanding of topological properties of quantum states had seen early success in explaining the conductivity of doped *trans*-polyacetylene (*trans*-PA) in terms of dispersionless soliton states [7,8]. These boundary states arise at the interface of the two degenerate bond-alternation patterns representing disparate topological electronic phases of PA [9]. In their seminal paper Su, Schrieffer and Heeger (SSH) described the occurrence of this exotic quantum states using a simple one-dimensional Tight Binding (TB) model [10,11]. Due to its exemplary role in describing chiral topological insulators, charge fractionalization and spin-charge separation in one-dimension (1D), there have been numerous efforts to realize the SSH-Hamiltonian in cold atom, photonic and acoustic experimental configurations [12-14]. To exploit the corresponding quantum states in electronic devices, it would however be desirable to rationally engineer topological electronic phases into stable and processable materials, a feat which has not been achieved so far. Here we present a flexible strategy to realize robust nanomaterials exhibiting valence electronic structures whose fundamental physics is described by the SSH-Hamiltonian. These solid-state materials are realized using atomically precise graphene nanoribbons (GNR) [19-21]. We demonstrate the controlled periodic coupling of topological boundary states [24] at junctions of armchair GNRs

of different widths to create quasi-1D trivial and non-trivial electronic quantum phases. Their topological class is experimentally determined by drawing upon the bulk-boundary correspondence [12] and measuring the presence (non-trivial) or absence (trivial) of localized end states by scanning tunneling spectroscopy (STS). The strategy we propose has the potential to tune the band width of the topological electronic bands close to the energy scale of proximity induced spin-orbit coupling [15] or superconductivity [16], and may allow the realization of Kitaev like Hamiltonians [3] and Majorana type end states [17].

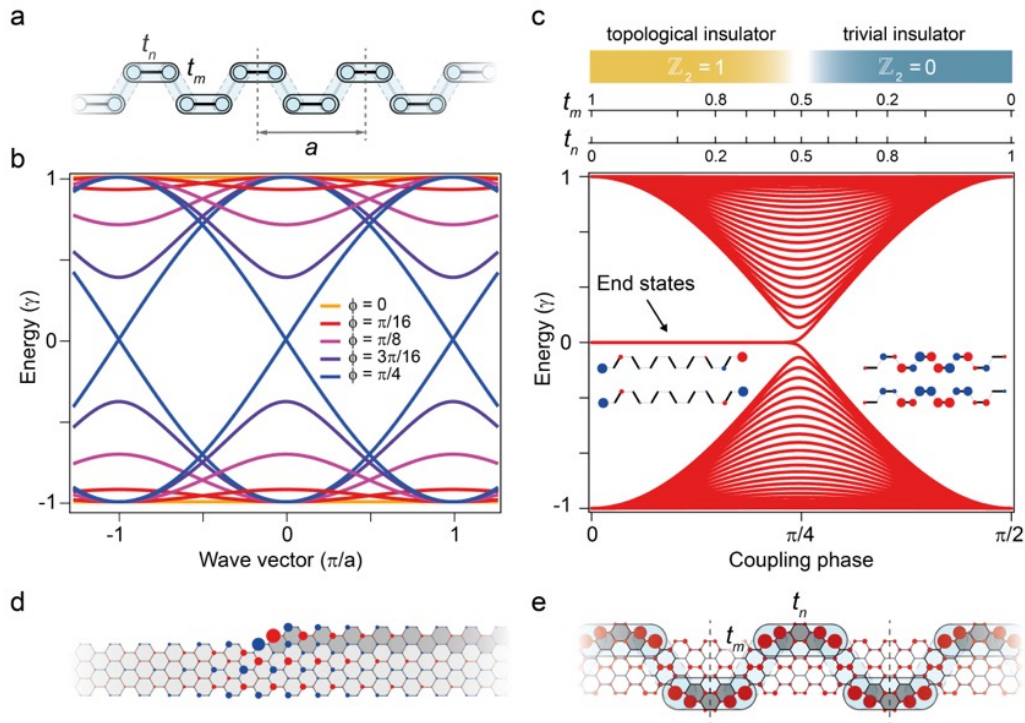

**Figure 1 | The SSH model and its realization in edge-extended graphene nanoribbons.**

**a**, Schematic representation of the dimerized $cis$-PA like SSH chain with illustration of the intra-cell coupling $t_n$ and inter-cell coupling $t_m$ within and between dimers, respectively. **b**, Dispersion relation $E(k)$ of the SSH chain displayed in (**a**) as a function of the phase factor $\phi$ which governs the coupling strengths $t_n$ and $t_m$ (see text). **c**, Energy level diagram as a function of $\phi$ for a finite SSH chain of 25 dimers, revealing topological zero energy modes for $\phi < \frac{\pi}{4}$ (i.e. $t_n < t_m$). The inset in (**c**) displays the wave functions of the frontier orbitals (i.e. closest to E=0) for a short 8 dimer chain with localized end state character for $\phi < \frac{\pi}{4}$ and extended bulk like character for $\phi > \frac{\pi}{4}$. **d**, Wave function of the $N$ to $(N + 2)$-AGNR boundary state at an isolated smooth 7-AGNR to 9-AGNR junction. Marker size in (**d**) denotes wave function amplitude and color (blue-red) parity. **e**, Schematic representation of the frontier orbitals (marker size denotes charge density) of a 7-AGNR with staggered edge extensions leading to short 9-AGNR segments. The corresponding 7- to 9-AGNR boundary states couple within the 9-AGNR segments ($t_n$) and across the 7-AGNR backbone ($t_m$), in analogy to the *cis*-SSH chain illustrated in (**a**).

The fundamental features of the SSH model which describes a 1D chain of dimerized, coupled and spinless fermion states are summarized in Fig. 1. Conceptually, the basic elements of the SSH model are an ensemble of equivalent fermion states $|\psi_i\rangle$ at each site $i$ of the chain, an intra-cell coupling between two such states within the same dimer denoted here by $t_n$, and an inter-cell coupling $t_m$ between states of neighboring dimers (Fig. 1a). The corresponding spinor based Hamiltonian $H(k) = d_x(k)\sigma_x + d_y(k)\sigma_y$ (with $d_x(k) = t_n + t_m \cos(k)$; $d_y(k) = t_n + t_m \sin(k)$) leads to the energy spectrum $E(k) = \pm\sqrt{t_n^2 + t_m^2 + 2t_n t_m \cos(k)}$ [11]. Depending on the relative strength of $t_n$ and $t_m$ three extremal phases can be distinguished: (*i*) An intra-cell decoupled, insulating phase with $E(k) = \pm t_m$ for $t_n = 0$ and $t_m \neq 0$; (*ii*) a metallic phase with $E(\pi) = 0$ and $E(0) = \pm 2t_n$ for equal coupling strengths $t_n = t_m \neq 0$; and (*iii*) an inter-cell decoupled, insulating phase with $E(k) = \pm t_n$ for $t_n \neq 0$ and $t_m = 0$.

Importantly, these three extremal solutions of the SSH chain can be smoothly connected by introducing a phase factor $\phi \in [0, \pi/2]$ governing the strength of $t_n$ and $t_m$ via $t_n = \gamma \cdot \sin^2(\phi)$ and $t_m = \gamma \cdot \cos^2(\phi)$, where $\gamma$ denotes the band width. The corresponding series of band structures (BS) $E(k, \phi)$ in Fig. 1b reveals non-dispersive BSs (orange) for the two insulating chain configurations at $\phi = 0$ and $\phi = \frac{\pi}{2}$, while for $\phi = \frac{\pi}{4}$ (blue) a gapless metallic phase is found. The fact that the smooth transition between two insulating phases can only occur by closing the gap is a clear evidence of their distinct topological class. This class can be assigned using the winding number $\vec{r}(k,\phi) = (d_x(k,\phi), d_y(k,\phi))$ around the origin as a $\mathbb{Z}_2$ topological invariant [11], which is $\mathbb{Z}_2 = 1$ for $\phi < \frac{\pi}{4}$ and $t_n < t_m$, making the corresponding phases topologically non-trivial, and topologically trivial with $\mathbb{Z}_2 = 0$ for $\phi > \frac{\pi}{4}$ and $t_n > t_m$. Note that, the winding number cannot be directly determined in experiments. Fortunately, the bulk-boundary correspondence, i.e. the relation between the bulk winding number and the existence or absence of boundary states, offers a convenient experimental approach to determine topological class. The discrete energy spectrum of a finite SSH chain of 25 dimers as a continuous function of $\phi$ is displayed in Fig. 1c. The topologically non-trivial phases for $\phi < \frac{\pi}{4}$ can readily be distinguished from the trivial ones with $\phi > \frac{\pi}{4}$ by the presence of two degenerate zero energy states localized at the left and right chain ends (insets of Fig. 1c).

Specifically designed graphene nanoribbons (GNRs) provide the platform to realize a novel class of robust solid state nanomaterials which can flexibly encompass all three aforementioned quantum phases of the SSH chain. The atomically precise structural control required to rationally engineer the corresponding electronic structures is achieved by on-surface synthesis [18]. Since the first successful bottom-up synthesis of GNRs by polymerization of dedicated molecular precursors [19], a wide variety of GNRs exhibiting different width, chirality, edge structure and chemical doping has been realized by means of on-surface synthesis [20-21]. In addition to the structurally flexible synthesis, the chemical robustness of GNRs allows handling under ambient conditions [22] and their integration into high performance electronic nanodevices [23]. This promises a more straightforward technological exploitation of GNR based topological quantum phases in applications than e.g. atom condensates in optical lattices [12].

The ability to flexibly engineer SSH-like topological quantum phases in GNRs firstly requires a suitable electronic state representing $|\psi_i\rangle$. We identify such a state in the zero energy boundary state present at the junction of two armchair graphene nanoribbons ($N$-AGNR) of different width. Here $N$ denotes the number of transverse carbon atom rows in the usual sense [19]. The boundary state we are considering here is itself of topological origin [24]. To understand this, we need to consider that $N$-AGNRs can be classified into three families according to their electronic properties. For $N = 3p$ and $N = 3p + 1$ ($p$ integer) the corresponding AGNRs exhibit a gapped electronic structure, whereas for $N = 3p + 2$ a gapless (i.e. metallic) behavior is observed at the TB level of theory [25]. At a smooth junction of a gapped $N$-AGNR with $N = 3p + 1$ and a gapped $N = 3p + 3$ AGNR (i.e., with two additional rows of carbon atoms) (Fig. S1-S4), a zero energy boundary state occurs due to the gapless $N = 3p + 2$ intermediate (Fig. S5-S9). This situation is analogous to PA, where the smooth transition from one bond alternation pattern to the complementary one can only proceed via closure of the gap and creates a zero energy soliton state [8,11]. The wave function of the corresponding boundary state at a 7-AGNR / 9-AGNR junction is displayed in Fig. 1d. Creating a periodic sequence of such boundary states along and across the $N$-AGNR backbone, by local extension to a finite $(N + 2)$-AGNR segment (Fig. 1e), produces an effective solid state analogue of a *cis*-SSH chain. The $(N + 2)$-AGNR segment length is denoted by the index $n$, and the separation across the backbone from one segment to the opposite one by $m$. The resulting staggered ($S$) ribbon structure is then labeled as $N$-AGNR-$S(n, m)$. Thereby, the structure shown in Fig. 1e with $N = 7, n = 1$ and $m = 3$ is denoted as 7-AGNR-$S(1,3)$ (see Fig. S1-S4 for details on the nomenclature). In terms of the SSH Hamiltonian, the segment length $n$ is directly related to the intra-cell coupling $t_n$ of the left and right boundary states, and the separation across the backbone denoted by $m$ determines the inter-cell coupling $t_m$.

The TB bulk BS (with $\gamma_0$=3 eV) of the staggered 7-AGNR-$S(1,3)$ is compared to the BS of the pristine 7-AGNR backbone in Fig. 2. The appearance of four dispersive bands around E=0 (corresponding to the Fermi energy $E_F$) in the band gap of the 7-AGNR backbone structure is readily observed (see Fig. S10-S11 for further $N$-AGNR-$S, I(n, m)$ examples). These bands can be fitted in excellent agreement with the zone-folded SSH energy spectrum $E(k)$ (blue solid lines in Fig. 2b) with $t_n = 0.45\ eV$ and $t_m = 0.59\ eV$ (or the other way around, due to the symmetry of $E(k)$ when exchanging $t_n$ and $t_m$).

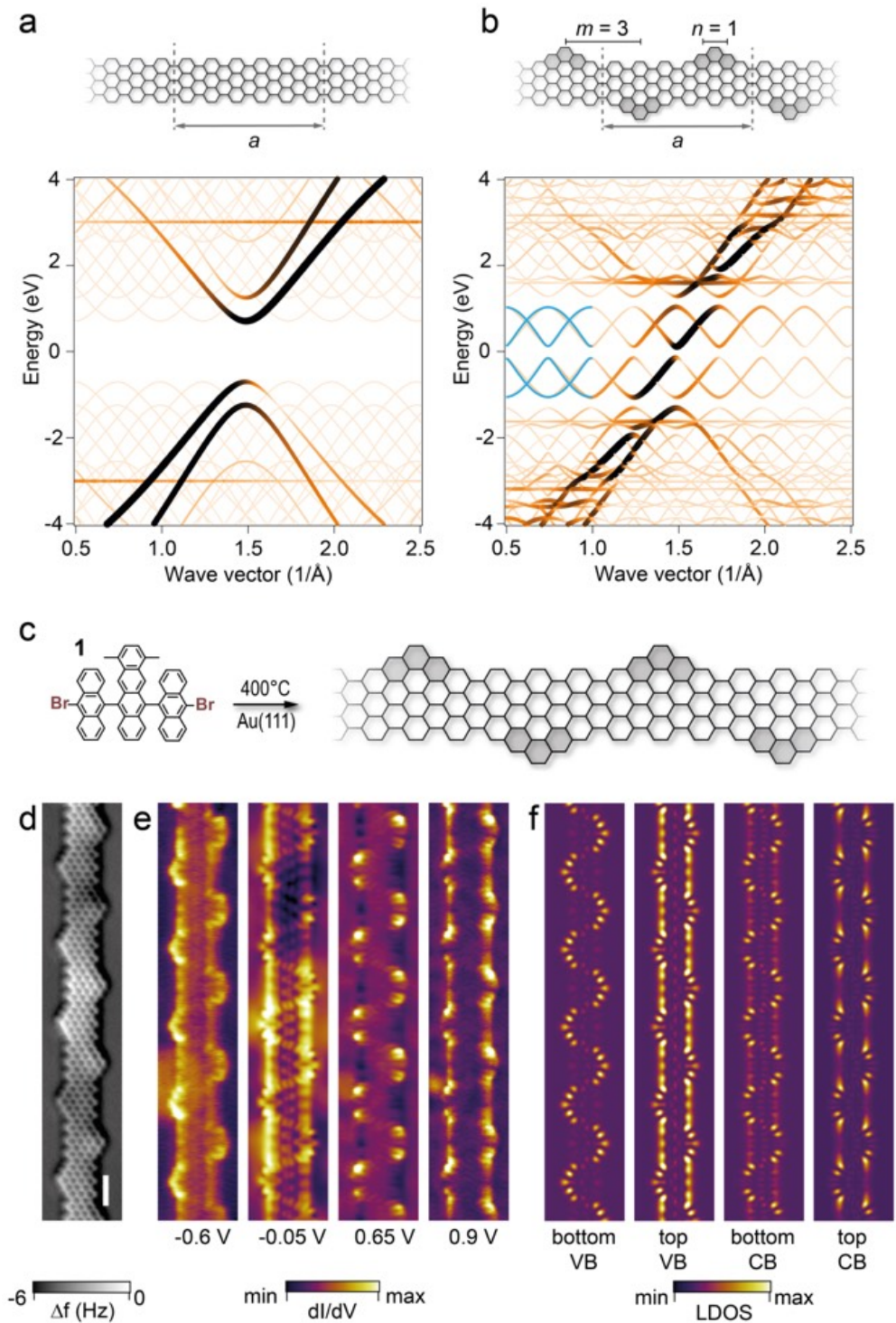

**Figure 2 | Electronic structure of the staggered edge-extended 7-AGNR-$S(1,3)$ nanoribbon.**

**a,b**, Structural models and TB band structure diagrams for the pristine 7-AGNR backbone (**a**) and the staggered 7-AGNR-$S(1,3)$ (**b**). Marker size and color (black to yellow) denote the magnitude of $|\langle \vec{k}|\psi_n(k)\rangle|$ i.e. the projection of the electronic states onto free electron states (Brillouin zone unfolding). Solid blue lines in (**b**) denote the analytical bands of the *cis*-SSH chain. **c**, Schematic representation of the on-surface synthesis route from monomer **1** to the 7-AGNR-$S(1,3)$. **d**, Constant-height nc-AFM image (with CO functionalized tip) of the frequency shift Δf of a 7-AGNR-$S(1,3)$ segment on Au(111). **e**, Series of constant-current dI/dV maps of the GNR shown in (**d**) at selected energies close to $E_F$ (which corresponds to 0 V sample bias). The set-point currents are 300 pA (U=-0.6 V), 600 pA (U=-0.05 V), 800 pA (U=0.65 V) and 1 nA (U=0.9 V). **f**, Sequence of TB derived constant-height charge density maps at the VB and CB extrema (at -1 eV, -0.2 eV, +0.2 and +1.0 eV from bottom VB to top CB). The 1 nm scale bar in (**e**) applies to all maps (**d**)-(**f**).

We have conceived a synthetic design to realize the staggered 7-AGNR-$S(1,3)$ structure by using 6,11-bis(10-bromoanthracen-9-yl)-1,4-dimethyltetracene (BADMT, monomer **1**) as precursor monomer. The methyl groups can form zigzag edges bridging smoothly the 7- and 9-AGNR segments through the cyclization with the neighboring aromatic rings, forming the intermediate 8-AGNR structure. The corresponding on-surface synthesis route (Fig. 2c) consists of the sublimation of monomer **1** onto a clean Au(111) surface, subsequent thermal precursor activation (dehalogenation) and polymerization at T=200 °C, and finally cyclodehydrogenation of the polymer at 400 °C. A constant-height non-contact atomic force microcopy (nc-AFM) image of the resulting structure is shown in Fig. 2d. The chemical stability of this GNR has been investigated by Raman spectroscopy (Fig. S22), and no spectral changes were detected after 5 days under ambient conditions. This observation is consistent with the reported high stability of the pristine backbone 7-AGNR [22].

STS investigation reveals that the 2.4 eV band gap of the pristine 7-AGNR on Au(111) [26,27] is drastically reduced to 0.65(10) eV for the 7-AGNR-$S(1,3)$. dI/dV maps of the main spectroscopic features around the gap (Fig. 2e) can be reliably assigned to bottom and top of the valence band (VB) and conduction band (CB), respectively, by comparison with TB simulations of the corresponding BS features (Fig. 2f). The experimentally observed $\Delta E_{exp}$=1.6 eV bandwidth is in good agreement with the one found from the TB calculations $\Delta E_{TB}$=$\sqrt{t_n^2 + t_m^2 + 2t_n t_m} = 2.08\ eV$ with $t_n = 0.45\ eV$ and $t_m = 0.59\ eV$. Because the symmetry of $E(k)$ with regard to exchange of $t_n$ and $t_m$ does not allow to determine which coupling term prevails, it remains open if the cis-PA SSH chain analogue realized with the 7-AGNR-$S(1,3)$ structure belongs to the topologically non-trivial class ($\mathbb{Z}_2 = 1$ with $t_m > t_n$) or the topologically trivial class ($\mathbb{Z}_2 = 0$ with $t_m < t_n$).

In order to clarify this we exploit the bulk-boundary correspondence [11] and check for the presence of end states at the termini of the $N$-AGNR-$S(n,m)$ nanoribbon family. There is, however, a complication arising from the concomitant presence of zigzag termini related end states of the $N$-AGNR backbone [28]. Both types of end states have topological origins but of different nature. As detailed in the Supplementary Materials (Fig. S16-S18), these two states can interact and hybridize such that the SSH end state is no longer present at zero energy. To prevent this, the terminus of the $N$-AGNR-$S(n,m)$ needs to be extended by a sufficiently long segment of the pristine $N$-AGNR backbone, as illustrated in Fig. 3a,b. This extension spatially separates the two topological states to minimize their interaction. Fig. 3a,b illustrates the resulting local density of states (LDOS) at the end of the $N$-AGNR-$S(n,m)$ segment (indicated by the arrows) as a function of $m$ for the 7-AGNR-$S(1,m)$ and 7-AGNR-$S(3,m)$ nanoribbon families, respectively.

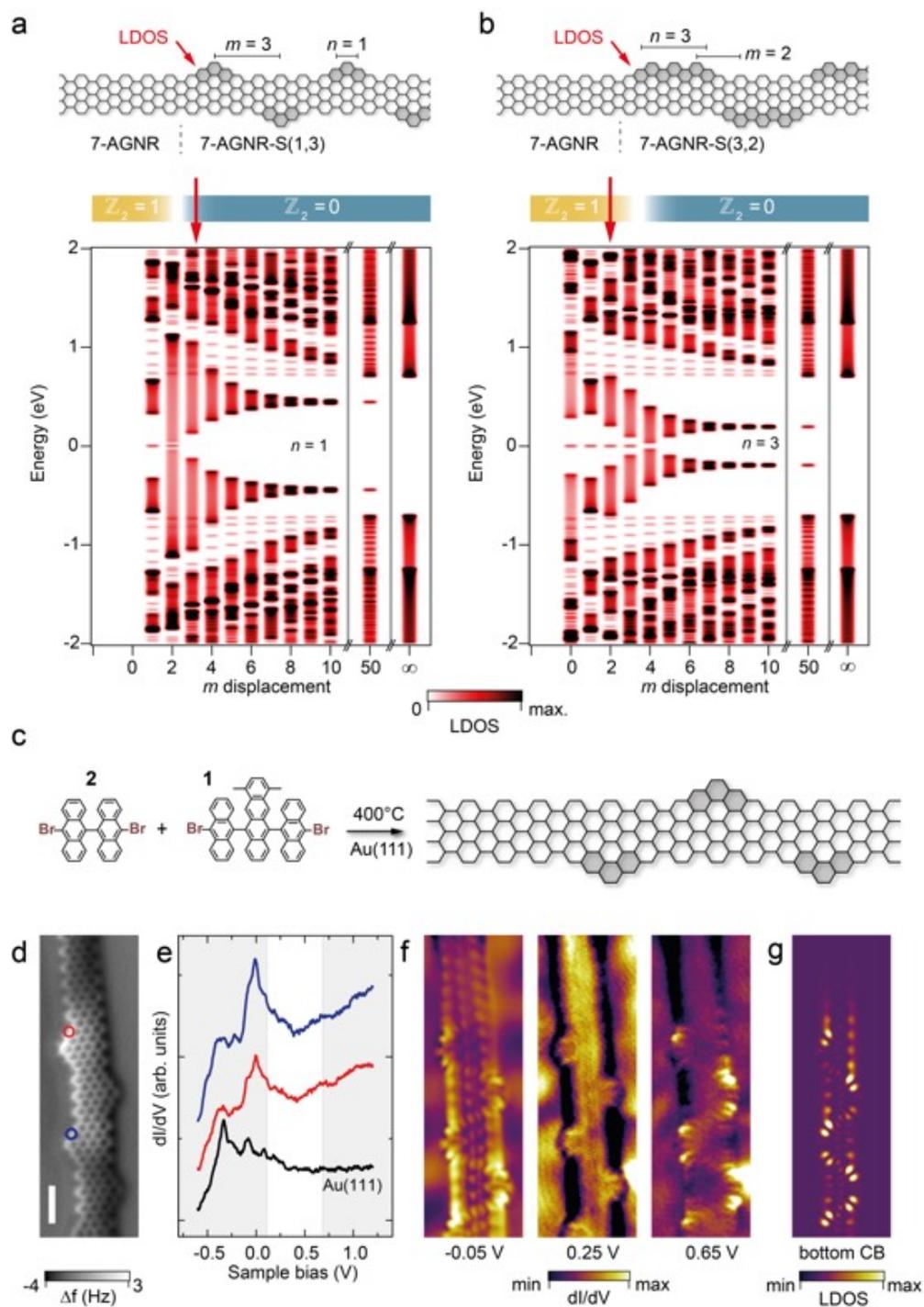

**Figure 3 | Bulk-boundary correspondence for the staggered edge-extended 7-AGNR-$S(n,m)$ nanoribbon family.**

**a,b**, LDOS plots for the 7-AGNR backbone extended 7-AGNR-$S(1,m)$ (**a**) and 7-AGNR-$S(3,m)$ (**b**) nanoribbon families, evaluated at the end of the 7-AGNR-$S(n,m)$ segment (see LDOS-labeled red arrows). The 7-AGNR-$S(1,3)$, whose structural model is depicted in (**a**), exhibits no zero energy end state (LDOS indicated by red arrow) and thus belongs to the topologically trivial $\mathbb{Z}_2 = 0$ class. Conversely, the 7-AGNR-$S(3,2)$ structure (model in (**b**), LDOS indicated by red arrow) reveals zero energy end states and thus belongs to the topologically non-trivial $\mathbb{Z}_2 = 1$ class. **c**, Synthetic pathway to the 7-AGNR backbone extended 7-AGNR-$S(1,3)$ nanoribbon using **1** and **2** as precursor molecules. **d**, Constant-height nc-AFM frequency shift (Δf) image of a 7-AGNR-$S(1,3)$ segment (acquired with CO functionalized tip). **e**, STS dI/dV spectra taken at positions indicated by the markers of the corresponding color in (**e**). **f**, Experimental constant-current dI/dV maps at the top of the VB (-0.05 V, I=200 pA), in the gap (+0.25 V, I=500 pA) and at the bottom of the CB (+0.65 V, I=500 pA) of the 7-AGNR-$S(1,3)$ shown in (**e**). **g**, TB simulated charge density map of the bottom of the CB, computed for the experimental structure (**d**).

The ($m = 1$) 7-AGNR-$S(1,1)$ exhibits a zero energy end state, indicating that it belongs to the topologically non-trivial phase ($\phi < \frac{\pi}{4}$) with $t_n < t_m$. Increasing $m$ decreases $t_m$ while $t_n$ remains approximately constant ($n = 1 = const.$). For $m = 2$ the LDOS shows a closing of the gap corresponding to $t_n \approx t_m$ ($\phi \approx \frac{\pi}{4}$), thus marking the metallic intermediate separating the non-trivial 7-AGNR-$S(1,1)$ from the trivial 7-AGNR-$S(1,3)$ which shows a gap again but with no zero energy end states. For $n = 3$ (Fig. 3b) $t_n$ is reduced and the non-trivial to trivial transition with $t_n \approx t_m$ should occur at larger $m$ (i.e. smaller $t_m$) compared to the $n = 1$ case. As can be seen from Fig. 3b, zero energy end states occur indeed for $m =$1, 2 and 3. This means that, according to the TB calculations, the experimentally realized 7-AGNR-$S(1,3)$ belongs to the topologically trivial $\mathbb{Z}_2 = 0$ class.

In order to verify this finding experimentally, the synthetic route shown in Fig. 2 was modified to allow the required extension of the staggered nanoribbon structure with a pristine 7-AGNR backbone segment. This is realized by sequential deposition of monomer **1** for the 7-AGNR-$S(1,3)$ and dibromo-bianthryl (DBBA, monomer **2**) for the 7-AGNR. Activation, polymerization, and cyclodehydrogenation are achieved by an appropriate annealing procedure as discussed before. The optimized protocol produces the required 7-AGNR – 7-AGNR-$S(1,3)$ heterostructures with good yield (Fig. 3d, Fig. S20). Differential conductance dI/dV spectroscopy at the end of the SSH GNR segment (red curve and marker in Fig. 3d,e) and at the internal SSH chain site (blue) shows nearly identical spectra and no indication of an end state. This is further corroborated by dI/dV mapping at selected energies around $E = 0$ (Fig. 3f). At U=-0.05 V the onset of the spatially extended VB states of the 7-AGNR-$S(1,3)$ can be seen, at U=0.25 V we are in the gap with no particular features, and at +0.65 V the bottom of the CB can be observed which is in good agreement with the TB charge density simulation of the lowest energy CB state of the 7-AGNR – 7-AGNR-$S(1,3)$ heterostructure (Fig. 3g). The experiment therefore confirms the TB prediction that the 7-AGNR-$S(1,3)$ is topologically trivial with $\mathbb{Z}_2 = 0$.

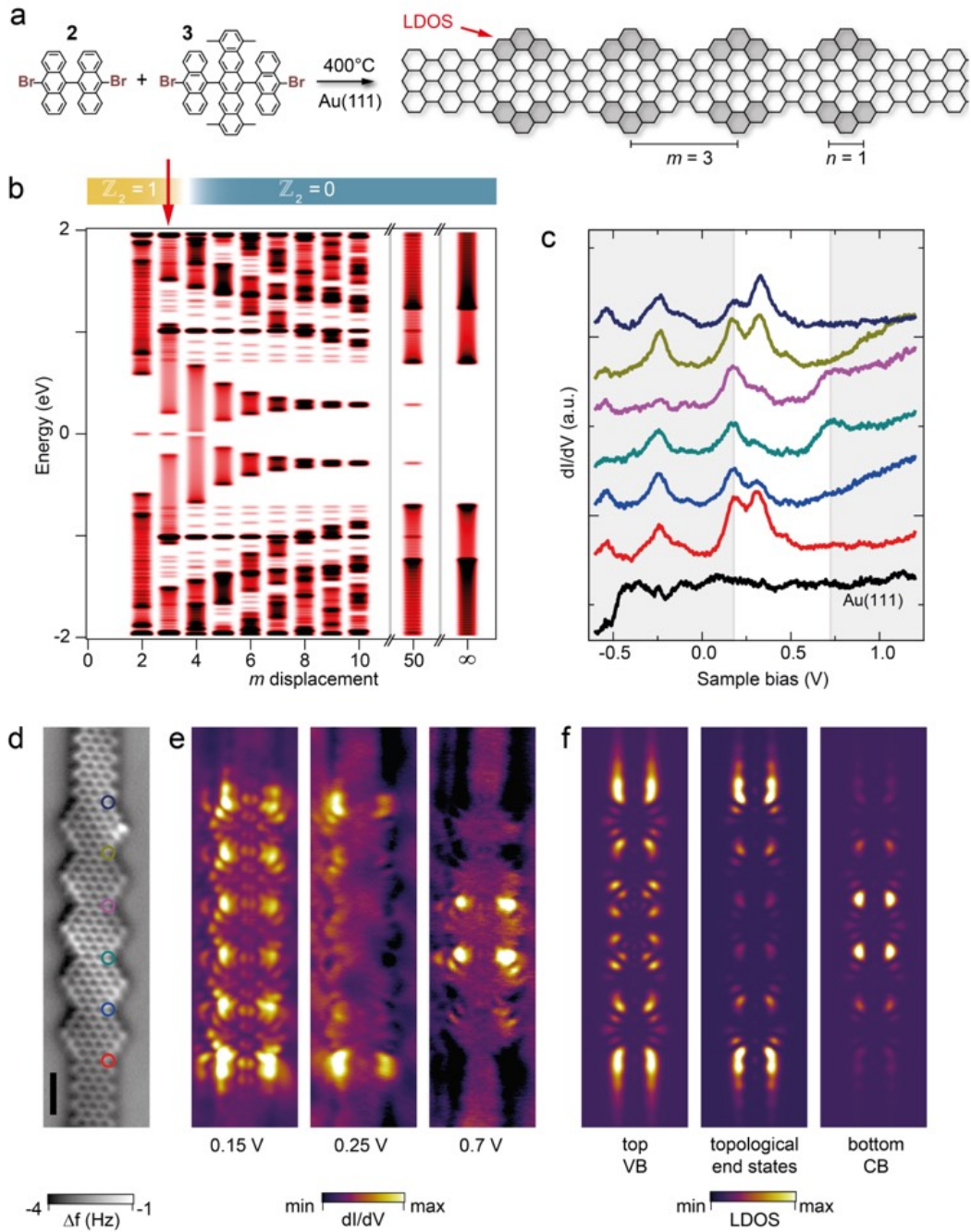

**Figure 4 | Non-trivial topological ($\mathbb{Z}_2 = 1$) phase of the inline edge-extended 7-AGNR-$I(1,3)$ structure. a**, On-surface synthesis route to the 7-AGNR backbone extended 7-AGNR-$I(1,3)$ nanoribbon. **b**, LDOS plots evaluated at the end of the 7-AGNR-$I(1,m)$ segment (see arrow in **a**) as a function of inter-segment spacing $m$, revealing the $\mathbb{Z}_2 = 1$ to $\mathbb{Z}_2 = 0$ transition at $m = 4$ with nearly complete gap closure and disappearance of the zero energy states for $m > 3$. **d**, Constant-height nc-AFM frequency shift ($\Delta f$) image of a 5 unit 7-AGNR-$I(1,3)$ segment with 7-AGNR extensions at both ends. **c**, dI/dV spectra (-0.6 V and 100 pA set-point before opening feedback loop) taken at the

locations indicated by the markers of corresponding color in (**d**). **e**, Experimental dI/dV maps of the main spectroscopic features at +0.15 V, +0.25 V and +0.7 V (all with I=500 pA). **f**, TB simulated charge density maps at the top of the VB, at E=0, and at the bottom of the CB, computed for the experimental structure (**d**).

As seen above, the staggered $N$-AGNR-$S(n,m)$ exhibits boundary states only for nanoribbon widths $N = 3p + 1$ ($p$ integer) and provides an electronic *cis*-PA analogue. If instead of an asymmetric $N$- to $(N + 2)$-AGNR junction an axially symmetric $N$- to $(N + 4)$-AGNR junction is considered, as illustrated in Fig. 4a, the resulting "inline" edge-extended GNR will yield zero energy boundary states for all backbone widths $N$ (Fig. S6). As for the staggered structure, the $(N + 4)$ segment length is denoted by $n$ and the segment spacing by $m$, and the inline "$I$" structure is thus labeled $N$-AGNR-$I(n,m)$. The structure shown in Fig. 4a is therefore a 7-AGNR-$I(1,3)$. The LDOS series for the 7-AGNR-$I(1,m)$ family in Fig. 4b reveals topological end states for $m$=2 and $m$=3 with the non-trivial to trivial phase transition between $m$=3 ($\mathbb{Z}_2 = 1$) and $m$=4 ($\mathbb{Z}_2 = 0$). The bulk BS for the 7-AGNR-$I(1,3)$ exhibits two *trans*-PA like bands with $t_n$=0.45 eV and $t_m$=0.65 eV (Fig. S11), which is very similar to the case of the staggered 7-AGNR-$S(1,3)$.

The synthetic route to the 7-AGNR extended 7-AGNR-$I(1,3)$ inline structure is analogous to the one for the staggered structure, but with 6,13-bis(10-bromoanthracen-9-yl)-1,4,8,11-tetramethylpentacene (BATMP, monomer **3**) used as the precursor molecule instead of monomer **1**. Fig. 4d presents the nc-AFM image of a 5 unit 7-AGNR-$I(1,3)$ that is extended by 7-AGNR segments at both ends. dI/dV spectra recorded at the 7-AGNR / 7-AGNR-$I(1,3)$ junction (dark blue and red in Fig. 4c and 4d) reveal a state at approximately 0.25 V that is only present at the chain ends. dI/dV mapping at this energy shows indeed that this state is spatially localized at the two ends of the 7-AGNR-$I(1,3)$ segment (Fig. 4e). Comparison with TB calculations (Fig. 4f) reveals that the extended state at 0.15 V can be assigned to the top of the VB, the 4 lobe state at +0.7 V in the center of the chain to the CB minimum, and that the state at +0.25 V is indeed the expected topologically non-trivial bulk-boundary end state (see Fig. S21 for a high resolution dI/dV map). This state is not observed at exactly 0 V due to charge doping of the nanoribbon by the substrate, which is well known to occur for low bandgap GNRs on Au(111) [29,30]. Altogether, this analysis shows that, in contrast to the staggered trivial 7-AGNR-$S(1,3)$, the inline edge-extended 7-AGNR-$I(1,3)$ belongs to the topologically non-trivial $\mathbb{Z}_2 = 1$ class and hosts topological end states.

The presence of short zigzag edge segments in the structure families discussed here suggests the possibility of magnetic ordering [31]. For the 7-AGNR-$S(1,3)$ and 7-AGNR-$I(1,3)$ structures the relatively strong coupling suppresses magnetic ordering, but the formation of antiferromagnetic spin-chains is expected for structures with larger $n$ and $m$ (Fig. S13-S15).

In summary, we have introduced the general concept of using topological zero energy states at the junctions of armchair graphene nanoribbon segments of different widths to construct robust solid state nanomaterials exhibiting valence electronic bands described by the SSH Hamiltonian. The topological class of the corresponding electronic configuration can be tuned by the spatial arrangement of these junctions. Furthermore, due to the insulator-metal-insulator transition occurring for different junction configurations of a given $N$-AGNR backbone, the band gap can be tuned over a wide range without changing the width of the GNR. Using dedicated on-surface synthesis routes we have grown edge-extended 7-AGNRs exhibiting topologically trivial and non-trivial quantum phases. Given their structural and chemical stability an integration of these topological nanomaterials in electronic devices is readily conceivable.

**Acknowledgements**

This work was supported by the Swiss National Science Foundation, the Office of Naval Research BRC Program, the Graphene Flagship (No. CNECT-ICT-604391), and the NCCR MARVEL. C.A.P. thanks the Swiss Supercomputing Center (CSCS) for computational support. X.Y. is grateful to the fellowship from the China Scholarship Council. O.G. thanks O. Yazyev and D. Bercioux for fruitful discussions.

**Author contributions**

O.G., P.R. and R.F. conceived and supervised this work. A.N., X.Y., X.F. and K.M. designed and synthesized the molecular precursors. S.W. performed the on-surface synthesis and SPM characterization. G.B.B. did the Raman analysis, C.D., A.C. and V.M. performed the corresponding simulations. C.A.P. did the DFT calculations. O.G. developed the conceptual framework, performed the TB calculations and wrote the manuscript, with contributions from all co-authors. P.R., S.W. and O.G. made the figures, with contributions from other co-authors.

**References**:

[1] Hasan, M. Z. & Kane, C. L. Colloquium : Topological insulators. *Rev. Mod. Phys.* **82**, 3045–3067 (2010)

[2] König, M. *et al.* Quantum Spin Hall Insulator State in HgTe Quantum Wells. *Science* **318**, 766–770 (2007)

[3] Kitaev, G. Unpaired Majorana fermions in quantum wires. *Physics-Uspekhi* **171**, 131–136 (2001)

[4] Bradlyn, B. *et al.* Topological quantum chemistry. *Nature* **547**, 298–305 (2017)

[5] de Vries, E. K. *et al.* Towards the understanding of the origin of charge-current-induced spin voltage signals in the topological insulator $Bi_2Se_3$. *Phys. Rev. B* **92**, 201102(R) (2015)

[6] Mourik, V. *et al.* Signatures of Majorana Fermions in Hybrid Superconductor-Semiconductor Nanowire Devices. *Science* **336**, 1003–1007 (2012)

[7] Chiang, C. K. *et al.* Electrical conductivity in doped polyacetylene. *Phys. Rev. Lett.* **39**, 1098 (1977)

[8] Su, W.-P., Schrieffer, J. R. & Heeger, A. J. Soliton excitations in polyacetylene. *Phys. Rev. B.* **22**, 2099–2111 (1980)

[9] Longuet-Higgins, H. C., F. R. S.& Salem, L. The alternation of bond lengths in long conjugated chain molecules. *Proc. R. Soc. Lond.* **A 251**, 172 (1959)

[10] Su, W., Schrieffer, J. R. & Heeger, A. J. Solitons in polyacetylene. *Phys. Rev. Lett.* **42**, 1698 (1979)

[11] Asbóth, J. K., Oroszlány, L., Pályi, O. A Short Course on Topological Insulators: Band-structure topology and edge states in one and two dimensions. *Lecture Notes in Physics* **Vol. 919** (Springer, Cham, 2016)

[12] Meier, E. J., An, F. A. & Gadway, B. Observation of the topological soliton state in the Su–Schrieffer–Heeger model. *Nat. Commun.* **7**, 13986 (2016)

[13] Tan, W., Sun, Y., Chen, H. & Shen, S.-Q. Photonic simulation of topological excitations in metamaterials. *Sci. Rep.* **4**, 3842 (2015)

[14] Chen, B. G. -g., Upadhyaya, N. & Vitelli, V. Nonlinear conduction via solitons in a topological mechanical insulator. *PNAS* **111**, 13004–13009 (2014)

**Materials and Methods**

**M1   Tight Binding calculations of electronic structure**

The calculations of the electronic structure are based on the nearest-neighbor hopping Tight Binding (TB) Hamiltonian considering the $2p_z$ orbital of the carbon atoms only:

$$H = \sum_i \varepsilon_i c_i^\dagger c_i - \sum_{<i,j>} \gamma_0 c_i^\dagger c_j \qquad (1)$$

Here $c_i^\dagger$ and $c_i$ denote the usual creation and annihilation operators on site $i$. $\langle i,j \rangle$ denotes the summation over nearest neighbor (NN) sites, the on-site energies $\varepsilon_i$ are all set to zero, and the NN hopping parameter is chosen to be $\gamma_0$=3 eV.

Band structures (BS) are calculated by taking into account the wave-vector dependent complex Bloch phase factors in the TB Hamiltonian. Unfolding of the BS into the extended Brillouin zone is achieved by projection of the wave functions of energy $E_n(\vec{k}_\parallel)$ on plane waves $|\langle \vec{k}_\parallel + \vec{k}_\perp | \psi_n(\vec{k}_\parallel) \rangle|$. The corresponding weight is displayed by marker size and color. Here $\vec{k}_\parallel$ and $\vec{k}_\perp$ denote the wave vectors parallel and perpendicular to the GNR axis, respectively. The perpendicular wave vector $\vec{k}_\perp$ for the projection is chosen non-zero in order to cut through the Dirac point of the parent graphene structure at $\vec{k}_\perp = 2\pi/3a$ and $\vec{k}_\parallel = 2\pi/\sqrt{3}a$ with a=2.44 Å being the length of the graphene basis vector.

Wave functions are reconstructed from the TB Eigenvectors $\alpha_{i,n}$ of energy $E_n$ by summing up the C $2p_z$ Slater type orbitals with $\xi$=1.625/$a_u$ over the atomic sites $i$ of the structure [1].

$$\psi_n(\vec{r}) = \sum_i \alpha_{i,n} \cdot z \cdot \exp(-\xi|\vec{r} - \vec{r}_i|) \qquad (2)$$

STS-mapping simulations are achieved in first approximation by displaying the charge density of the considered states in the energy interval $[\varepsilon_1, \varepsilon_2]$ at constant height $z_0$ according to:

$$LDOS(x, y, z_0) = \sum_n |\psi_n(\vec{r})|^2 \qquad \text{for all } n \text{ with } E_n \in [\varepsilon_1, \varepsilon_2] \qquad (3)$$

The results of the BS calculations of the 7-AGNR-S(1,3) and 7-AGNR-I(1,3) structures are compared to DFT calculations in Fig. S12.

**M2   Molecular precursor and nanoribbon synthesis**

The chemical synthesis of the monomers **1** (BADMT), **2** (BATMP) and **3** (DBBA) is detailed in the Supplementary Materials section together with details of the on-surface synthesis of the corresponding GNRs (Scheme S1, Fig. S23-S33, Fig. S19-S22).

**M3   STM/STS and nc-AFM characterization**

A commercial low-temperature STM/AFM system (Scienta Omicron) with a base pressure below $1\times10^{-10}$ mbar was used for sample preparation and characterization under ultra-high vacuum conditions. Au(111) single crystal substrates were cleaned by standard argon sputtering and annealing cycles. Molecular precursors were deposited from a homemade six-fold evaporator, which allows sublimation of two molecular species at the same time. The deposition flux was around 2 Å/min as determined by a quartz crystal microbalance. STM images and differential conductance dI/dV maps were recorded in constant-current mode unless noted otherwise. Constant-height tunneling current and nc-AFM frequency shift images were recorded with a CO-functionalized tip attached to a quartz tuning fork sensor (resonance frequency 23.5 kHz). dI/dV spectra were recorded using the lock-in technique ($U_{rms}$ = 20 mV at 680 Hz modulation). All data shown were acquired at a sample temperature of 5 K.

# Supplementary Information

# Engineering of robust topological quantum phases
# in graphene nanoribbons

Oliver Gröning, Shiyong Wang, Xuelin Yao, Carlo A. Pignedoli, Gabriela Borin Barin, Colin Daniels, Andrew Cupo, Vincent Meunier, Xinliang Feng, Akimitsu Narita, Klaus Müllen, Pascal Ruffieux & Roman Fasel

## 1. Structure classification and nomenclature

The atomic structure of the graphene nanoribbons (GNRs) that we discuss in this work is based on an armchair GNR (AGNR) backbone having $N$ rows of carbon atoms in the transverse direction ($N$-AGNR). This $N$-AGNR backbone is locally asymmetrically extended to a $(N + 2)$-AGNR or symmetrically extended to a $(N + 4)$-AGNR, as illustrated in Fig. S1. These edge extensions produce boundary regions at the junction to the $N$-AGNR backbone.

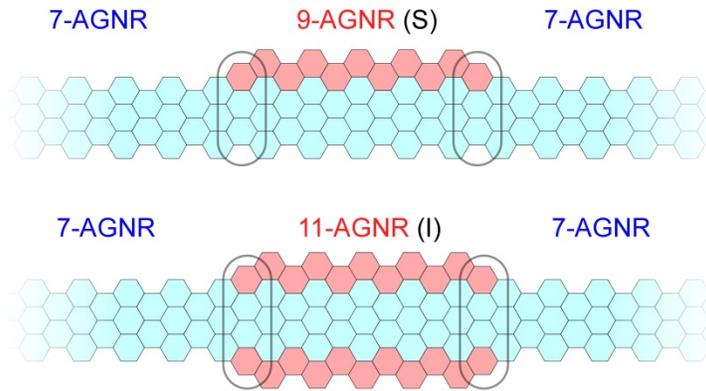

**Figure S1**
Schematic representation of the structural elements of the $(N + 2)$-AGNR (top) and $(N + 4)$-AGNR (bottom) segments on the $N$-AGNR backbone. The displayed structures are exemplary for a $N = 7$ backbone (7-AGNR). The left and right junction regions between the 7-AGNR backbone and the 9- or 11-AGNR segments, respectively, are highlighted by the ovals.

The periodic arrangement of these basic segments on the $N$-AGNR backbone yields non-trivial, edge extended AGNR structures. The two specific structures of interest in the context of this work are the so-called "staggered" ($S$-type, based on asymmetric $N + 2$ segments) and the "inline" ($I$-type, based on axially symmetric $N + 4$ segments) structures that are discussed in detail here below.

## 1.1. The staggered ("$S$"-type) structure

For the staggered $S$-structure the segment length is denoted by the integer $n \geq 1$, which refers to the number of units constituting the $(N + 2)$-AGNR segment (Fig. S2a).

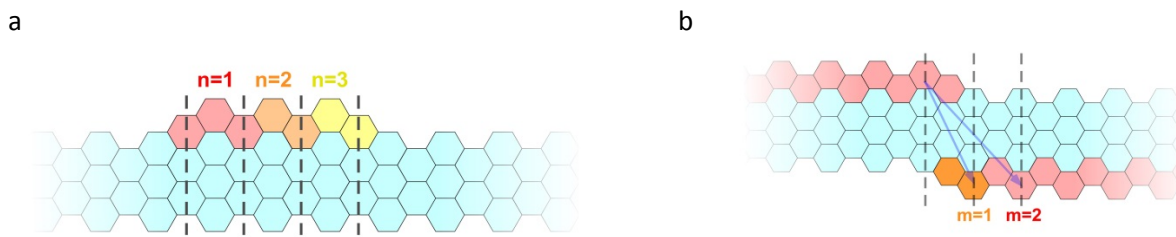

**Figure S2**
**a**, Schematics of edge extension producing a $(N + 2)$-AGNR segment structure (reddish to yellow color) on a $N$-AGNR backbone (light blue color). The length of the $(N + 2)$-AGNR segment is denoted by the index $n$. **b**, Schematic illustration of the inter-edge separation $m$ of two $(N + 2)$ segments.

In order to realize the periodically staggered structure of interest, the edge extensions producing the $(N + 2)$ segments are alternatingly placed at opposite edges of the $N$-AGNR backbone. The separation of the segments is denoted by the index $m$, which denotes the axial distance of the left hand terminus of the lower edge segment with respect to the right hand terminus of the upper edge segment (Fig. S2b). The same separation applies for the left hand terminus of the upper edge segment with respect to the right hand terminus of the lower edge segment.
The index $m$ denotes the axial displacement in terms of unit cells of the backbone ($a_0=3*a_C=3*1.41$ Å=4.23 Å). Although $m$ can be smaller than 1 we do not consider such structures because the resulting segment overlap will yield local $(N + 4)$-AGNR structures.

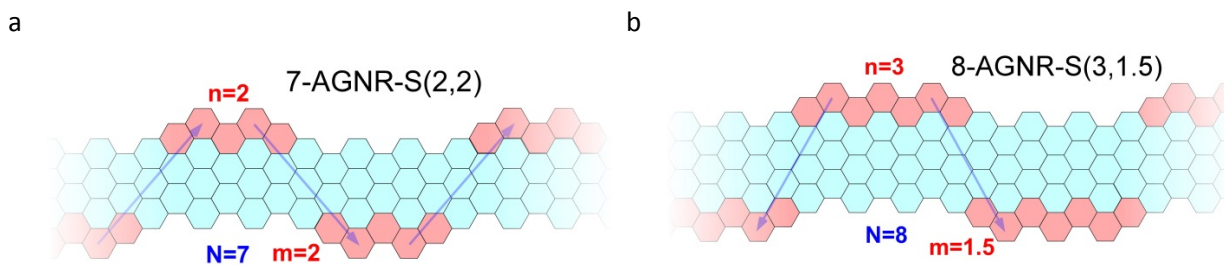

**Figure S3**
Examples of staggered edge-extended AGNRS: **a**, $N = 7$, $n = 2$ and $m = 2$ ("7-AGNR-$S(2,2)$"); and **b**, $N = 8$, $n = 3$ and $m = 1.5$ ("8-AGNR-$S(3,1.5)$").

The resulting staggered edge-extended GNR structures are denoted as $N$-AGNR-$S(n, m)$, with:

$N$:    Width of the backbone, counting the number of transverse rows of carbon atoms.
$S$:    An identifier for "**S**taggered"
$n$:    $(N + 2)$ segment length, starting with $n \geq 1$
$m$:    Across the ribbon $(N + 2)$ segment spacing, with $m \geq 1$ ($N$ odd $\rightarrow m$ integer, $N$ even $\rightarrow m$ half integer)

## 1.2. The inline ("$I$"-type structure

In addition to the staggered ($S$) structure the inline ($I$) arrangement of two ($N + 2$) aligned segments across the $N$-AGNR backbone is possible. $N$, $n$ and $m$ retain their meaning with regard to the $S$-structure with the index $m$ denoting the separation of the segments along the same edge as well as across the backbone. In contrast to the $S$-structure, for the $I$-structure the $N$-AGNR backbone is locally extended to a finite length ($N + 4$)-AGNR segment with $N$ being odd.

The nomenclature for this inline arrangement is $N$-AGNR-$I(n, m)$, with:

$N$: Number of transverse rows of carbon atoms in the backbone $N$, must be odd.
$I$: An identifier for "**I**"nline"
$n$: ($N + 4$) segment length, starting with ($n \geq 1$)
$m$: Across the ribbon ($N + 4$) segment spacing ($m \geq 2$)

Figure S4 shows schematically a 7-AGNR-$I(2,3)$ (**a**) and a 7-AGNR-$I(1,4)$ structure (**b**).

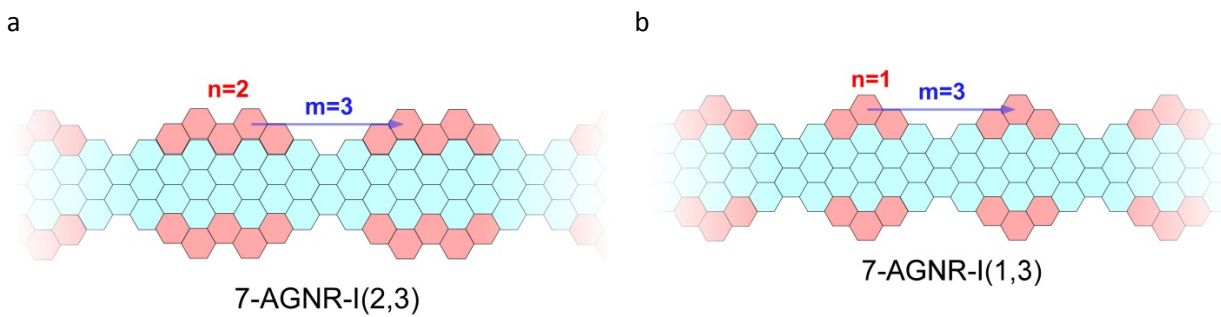

**Figure S4**
Examples of inline edge-extended AGNRs: **a**, 7-AGNR-$I(2,3)$, and **b**, 7-AGNR-$I(1,3)$.

## 2. Electronic properties of the $S$ and $I$ edge-extended AGNR structures

The specific electronic properties of the $S$- and $I$-structures are rooted in the particular electronic character of their parent $N$-AGNR.

### 2.1. The $N$-AGNR backbone structure

The $N$-AGNRs can be classified according to their electronic character in 3 families [1]: Two semiconducting families with $N = 3p$ or $N = 3p + 1$ and a metallic family with $N = 3p + 2$, with $p \geq 0$ integer. These families can be well distinguished by the evolution of their energy gap with increasing ribbon width $N$, as displayed in Fig. S5 (from Tight Binding (TB) calculations with the nearest neighbor hopping $\gamma_0$=3 eV).

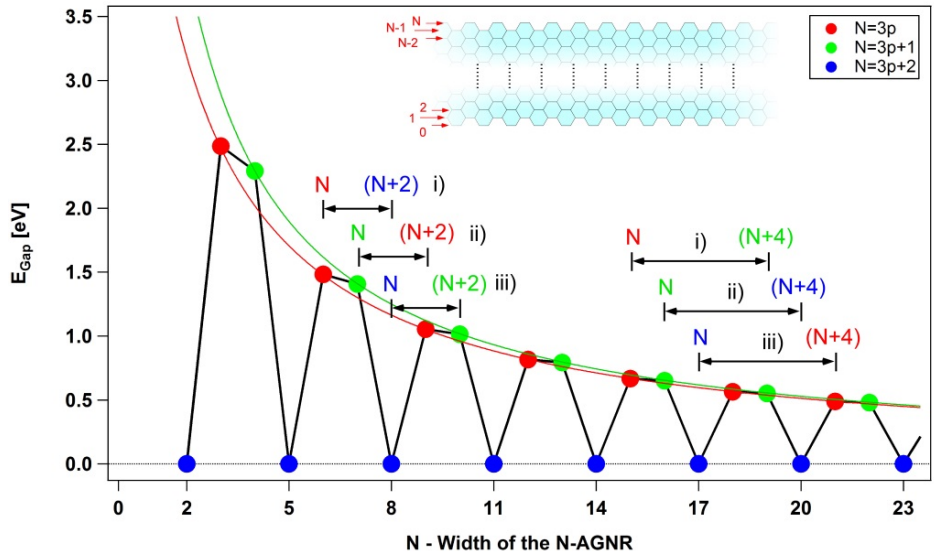

**Figure S5**

Dependence of the $N$-AGNR nearest neighbor TB energy gap on ribbon width $N$. The electronic family is denoted by the marker color. The three electronically different types of junctions i) – iii) for the staggered $N/(N + 2)$ (Fig. S3) and the inline $N/(N + 4)$ (Fig. S4) structures are indicated.

The sequence of semiconducting and metallic properties of AGNRs as a function of ribbon width $N$ can be regarded as a discrete topological phase diagram, where the metallic structures represent the boundaries separating one topological family of semiconducting $N$-AGNRs (e.g. $N = 3p$ or $N = 3p + 1$) from the next one (e.g. $N = 3p^*$ or $N = 3p^* + 1$ with $p^* = p + 1$). In this context the junctions formed between the $N$-AGNR backbone and the edge-extended $N + 2$ segments of the $S$-structure (or the $N + 4$ segments of the $I$-structure) can be regarded as "smooth" electronic transition regions within the same topological class if the transition can proceed without closing the gap (i.e. there is no intermediate metallic $N$-AGNR). On the other hand, the transition is considered to be across topological classes if it cannot proceed without closing the gap, i.e. if there is a metallic $N$-AGNR intermediate. Although a gap closure is not always required it is a hallmark of a topological phase transition [2]. The topological nature of AGNRs and the occurrence of boundary states have been discussed recently by Louie et al. [3].

As we will see, this has consequences on the appearance of topologically protected zero energy (E=0) boundary states, as discussed in the following.

## 2.2. Boundary states at the $S$- and $I$-segment junctions and their on-segment interaction

Considering a $N$-AGNR backbone with a single edge-extended segment (as shown in Fig. S1) one realizes that the junctions between the $N$-AGNR backbone and the $(N+2)$- respectively $(N+4)$-AGNR segments mark the transition from one electronic structure type to another. Accordingly, as a function of the width $N$ of the backbone $N$-AGNR we can identify three different junction types:

For the staggered edge extensions ($S$-structure):

i)     $N = 3p$     (sc)    →    $(N+2) = 3p+2$     (met)
ii)    $N = 3p+1$     (sc)    →    $(N+2) = 3p+3 = 3p^*$     (sc with $p^* = p+1$)
iii)   $N = 3p+2$     (met)    →    $(N+2) = 3p+4 = 3p^*+1$     (sc with $p^* = p+1$)

For the inline edge extensions ($I$ structure):

i)     $N = 3p$     (sc)    →    $N+4 = 3p+4 = 3p^*+1$     (sc with $p^* = p+1$)
ii)    $N = 3p+1$     (sc)    →    $N+4 = 3p+5 = 3p^*+2$     (met with $p^* = p+1$)
iii)   $N = 3p+2$     (met)    →    $N+4 = 3p+6 = 3p^*$     (sc with $p^* = p+2$)

where "*sc*" and "*met*" indicate the semiconducting and metallic nature of the corresponding ribbon structure, respectively.

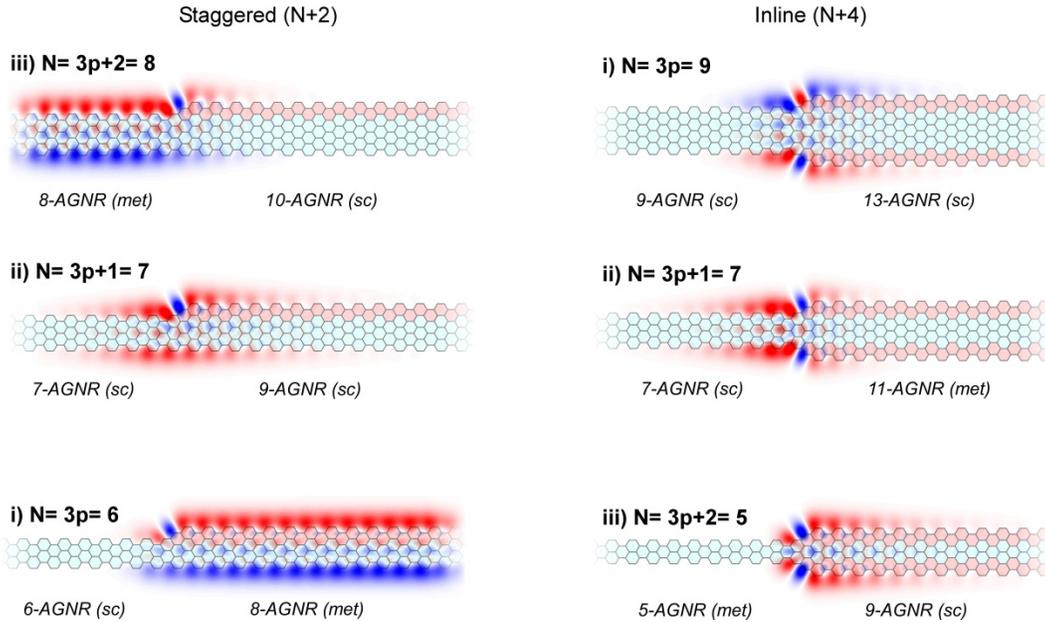

**Figure S6**
Catalogue of highest occupied molecular orbital (HOMO, i.e. closed to E=0) TB wave functions for the six possible junction types i) to iii) for the staggered $N \to (N+2)$-AGNR and the inline $N \to (N+4)$-AGNR junctions.

It can be seen from Fig. S6 that the different junction types yield different types of low energy states. In the staggered case only junction type ii) with $N = 3p + 1$ yields a zero-energy state localized at the junction. In the two other cases i) and iii) the states closest to E=0 are the metallic states of either the backbone (for $N = 3p + 2$) or of the metallic $(N + 2)$-segment (for $N = 3p$). In contrast the inline $(N + 4)$ structure exhibits localized junction states for all three cases i)-iii).

If the segment length $n$ is reduced to finite values, the junction states at the left and right ends of the edge-extended segment can no longer be regarded as isolated because they can interact with each other. In fact, a segment of finite length $n$ can be viewed as the intra-cell dimer of the SSH model. Depending on segment length $n$, the boundary states at either end (left $|L\rangle$ and right $|R\rangle$) will hybridize to form bonding and antibonding hybrid states with energy $E = \pm t_n$ where $t_n = \langle L|H|R \rangle$ ($H$ denoting the Hamiltonian of the system). The energy splitting $t_n$ will generally decrease (exponentially for large $n$) with $n$ as shown in Fig. S7.

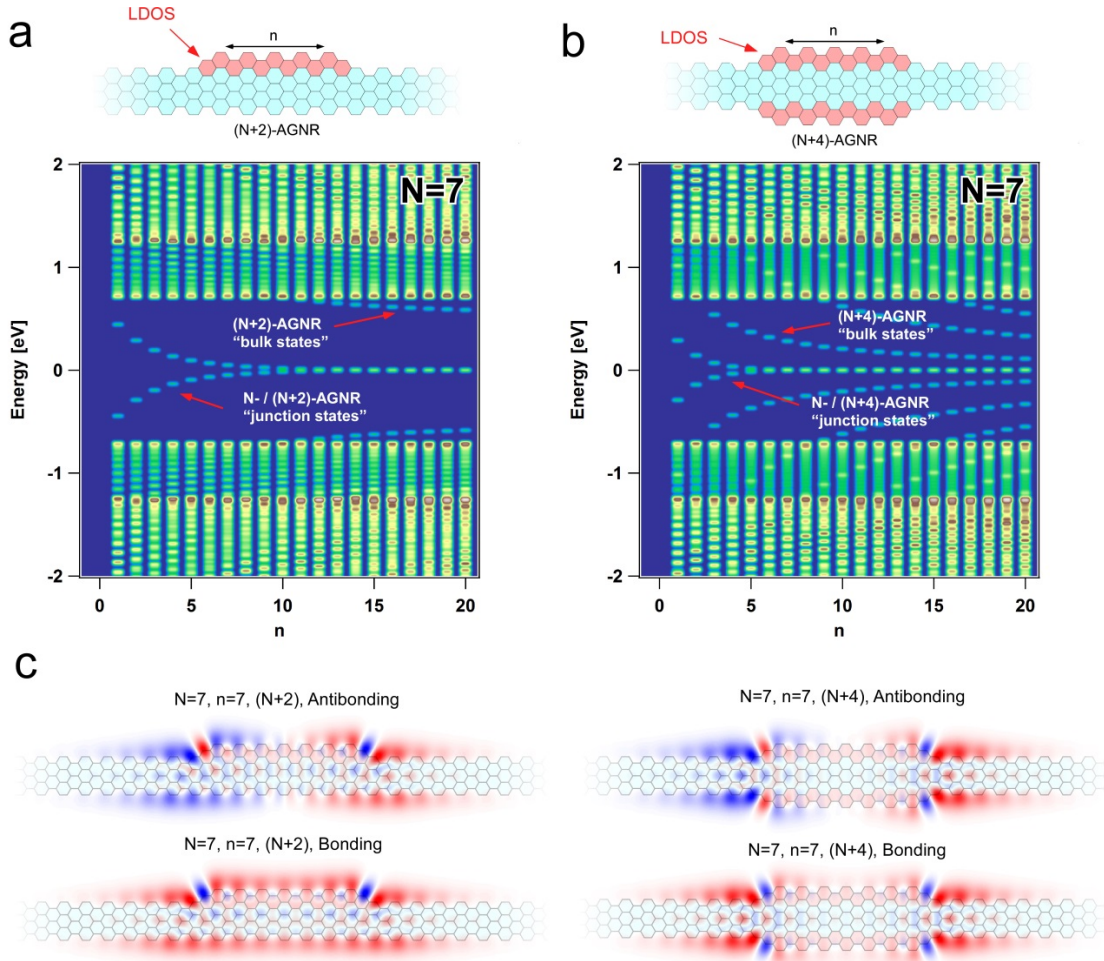

**Figure S7**
Local density of states (LDOS) maps (DOS at the segment boundary) for **a**, an isolated 9-AGNR and **b**, an isolated 11-AGNR segment of varying length $n$ within a 7-AGNR backbone. **c**, Wave function plots for the bonding and antibonding hybrid states for $n$=7 in the two cases.

The interaction of the $|L\rangle$ (left) and $|R\rangle$ (right) boundary states does not only depend on segment length $n$ but also on backbone width $N$ as can be seen from the LDOS maps of Fig. S8.

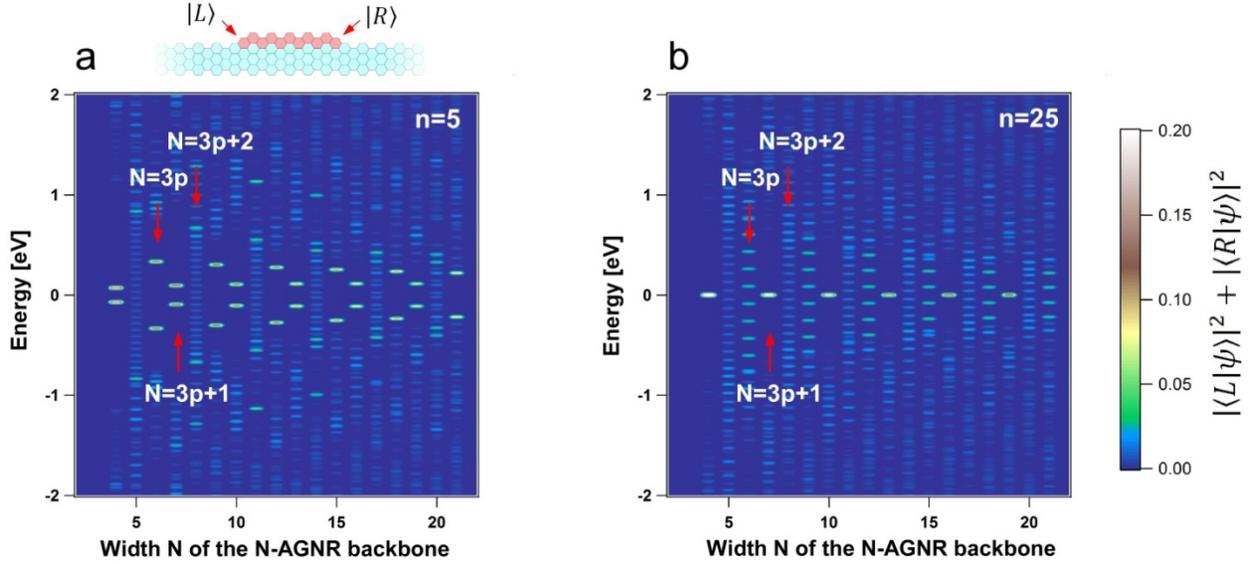

**Figure S8**
LDOS maps (DOS at the segment boundaries) for an isolated $(N+2)$-AGNR segment within a $N$-AGNR backbone for backbone widths $N = 4, 5, \ldots, 21$. **a**, Short segment length $n$=5. **b**, Long segment length $n$=25.

The three junction types i) - iii) discussed above can be well distinguished in the energy level spectra for different $N$ (Fig. S8a). For the case iii) with $N = 3p + 2$ the $N$-AGNR backbone is metallic and no states with particularly strong weight at the junction can be discerned. The situation is different for the cases i) with $N = 3p$ and ii) with $N = 3p + 1$, where two low-energy states localized in the junction region (i.e. with high amplitude in the LDOS) can clearly be observed. If the $N + 2$ segment length is increased from $n = 5$ to $n = 25$ (Fig. S8b), localized states remain only in the case ii) with $N = 3p + 1$. In the limit of $n \to \infty$ this junction state collapses to $E \cong 0$ with 2-fold degeneracy.

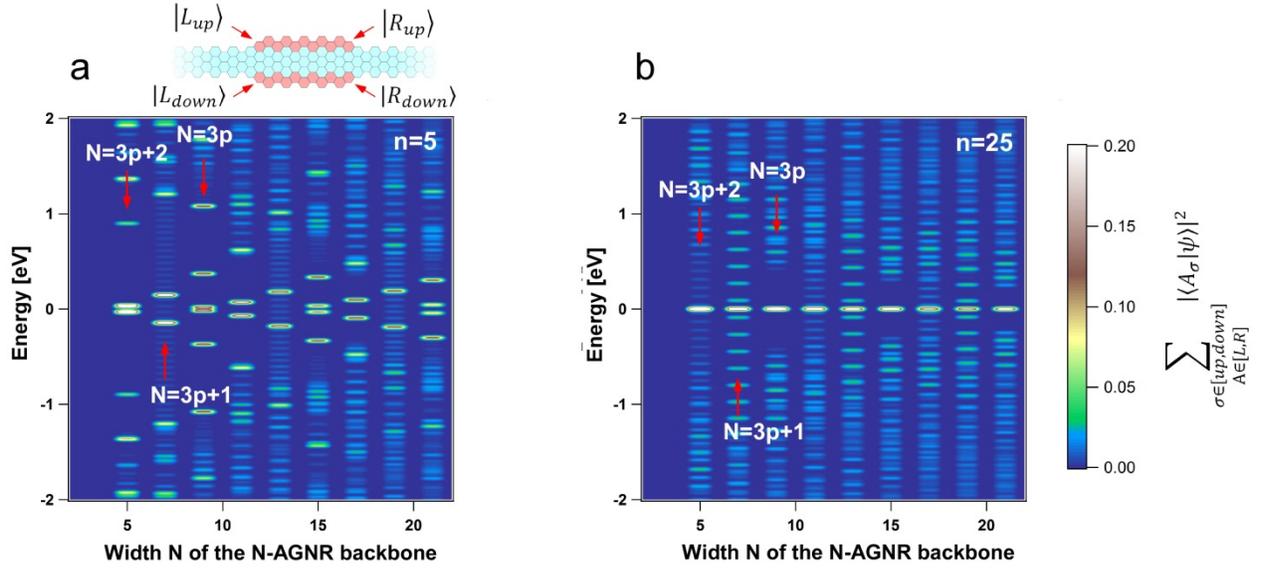

**Figure S9**

LDOS maps (DOS at the segment boundaries) for an isolated $(N+4)$-AGNR segment within a $N$-AGNR backbone for backbone widths $N = 5, 7, \ldots, 21$. **a**, Short segment length $n=5$. **b**, Long segment length $n=25$. The color scale displays the summed square of the wave function amplitude on the center atoms of the four zigzag kink sites.

In the case of the inline edge-extended $(N+4)$-AGNR segment again all three cases i) - iii) can be well distinguished in the sequence of energy level spectra for increasing backbone width $N$ (Fig. S9). For the short segment length $n = 5$, states localized at the junctions are observed in all cases. If the segment length is increased to $n = 25$ (Fig. S9b) a 2-fold degenerate zero-energy state appears for all backbone widths $N$, and not just for $N = 3p + 1$ as in the case of the $S$-type structure.

This behavior can be understood when looking at the staggered $N$-AGNR / $(N+2)$-AGNR and the inline $N$-AGNR / $(N+4)$-AGNR junctions as regions of gradual increase of AGNR width, i.e. by considering the $(N+1)$ and the $(N+1)$, $(N+2)$ and $(N+3)$ intermediate widths for the two cases, respectively (Fig. S6). For the staggered structure only the $N = 3p+1$ case (case ii) in Fig. S6) features a metallic $(N+1) = 3p+2$ intermediate. In this case we can understand the junction as connecting two insulating/semiconducting AGNR segments via closure of the gap. Accordingly, we have the appearance of a topologically protected zero-energy state at the junction, very much like for the soliton case of polyacetlylene [4]. For the inline structure there is always an intermediate width with metallic properties between the $N$-AGNR and the $(N+4)$-AGNR segments, and we will thus find a topologically protected zero-energy state at the junction for all AGNR backbone widths $N$.

## 2.3. Periodic coupling of junction states

For small $n$ and $m$ the junction states interact with each other and split in energy. Furthermore, the junction states can hybridize with the metallic $N$-AGNR backbone for case iii) or the metallic $(N+2)$ segment in case i) of the staggered $N$-AGNR-$S(n,m)$.

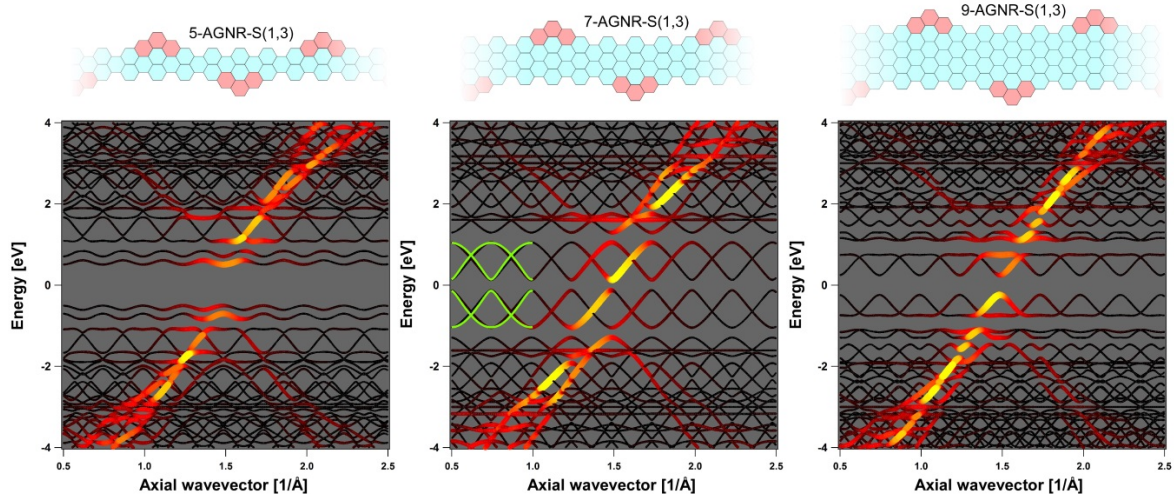

**Figure S10**
Band structure calculations for the staggered edge-extended 5-AGNR-$S(1,3)$ (left), 7-AGNR-$S(1,3)$ (middle) and 9-AGNR-$S(1,3)$ (right), representing the cases iii), ii) and i), respectively. The green solid lines in the panel 7-AGNR-S(1,3) denotes the fit with the analytical solution of the $cis$-PA structure.

The impact of hybridization can clearly be observed in Fig. S10. Only the case where $N = 7$ (i.e. $N = 3p + 1$ with $p = 2$) shows nearly undistorted $cis$-PA like bands around E=0. In the case $N = 5$ (i.e. $N = 3p + 2$) no such band can be identified, and for $N = 9$ (i.e. $N = 3p$ with $p = 3$) the PA like bands hybridize at about +/- 0.8 eV with the valence and conduction band states of the backbone AGNR.

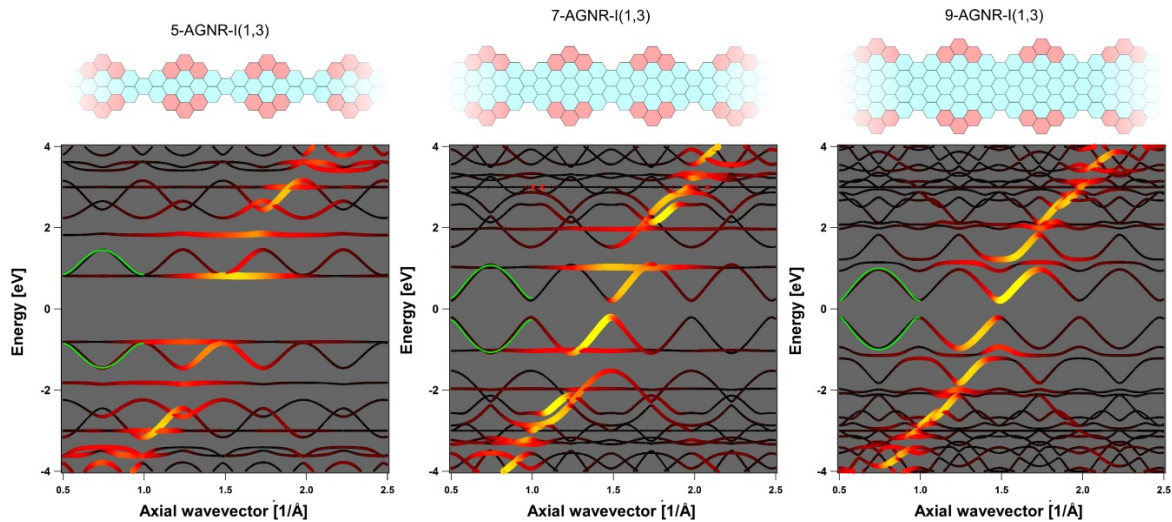

**Figure S11**
Band structure calculations for the inline edge-extended 5-AGNR-$I(1,3)$ (left), 7-AGNR-$I(1,3)$ (middle) and 9-AGNR-$I(1,3)$ (right), representing the cases iii), ii) and i), respectively. As in Fig. S10 the solid green lines denote the fits with the analytical solution for the $trans$-PA structure.

For the $I$-type structures shown in Fig. S11 the $trans$-PA like bands (highlighted in light green) near E=0 show little tendency of hybridization.

## 2.4. Comparison of Tight Binding (TB) and Density Functional Theory (DFT) band structures

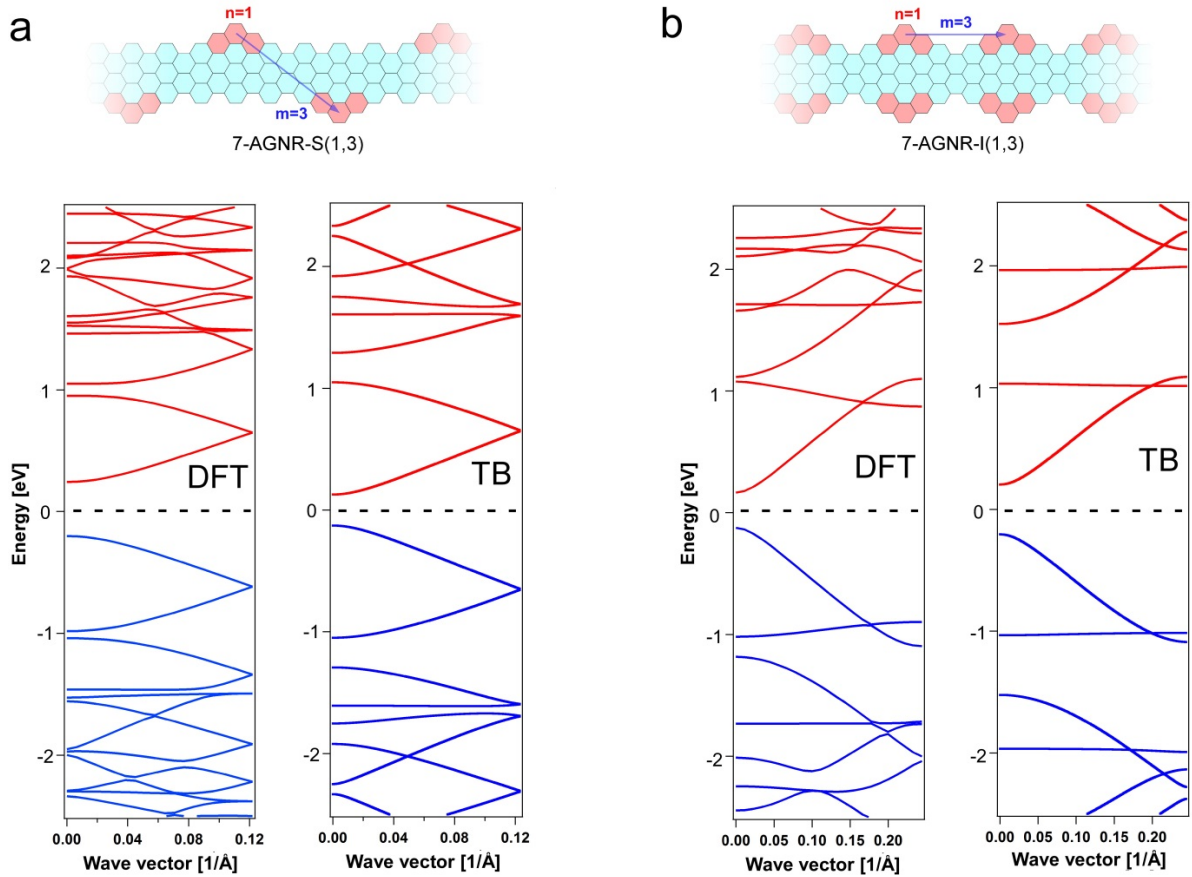

**Figure S12**

Comparison of the DFT bands close to $E_F$ (dotted line) with the nearest neighbor TB bands calculated with $\gamma_0$=3 eV. **a**, The staggered edge-extended 7-AGNR-$S(1,3)$. **b**, The inline edge-extended 7-AGNR-$I(1,3)$.

In Fig. S12 we compare the band structures obtained from TB with the ones from DFT. Very good agreement is observed for the frontier bands around the Fermi energy $E_F$ (dashed line).

The DFT simulations were performed with the plane wave code quantum espresso [5]. The generalized gradient approximation within the PBE [6] parameterization was used for the exchange correlation functional. Ultrasoft pseudopotentials from the Standard Solid State Pseudopotentials accuracy library [7] were employed to model the ionic potentials. A cutoff of 50 Ry (400 Ry) was used for the plane wave expansion of the wave function (charge density). For the convergence of the wave function a grid of 10 / 17 k-points was used to sample the 1D Brillouin zone for the 7-AGNR-$S(1,3)$ / 7-AGNR-$I(1,3)$ ribbons, respectively. The unit cell of the 7-AGNR-$S(1,3)$ / 7-AGNR-$I(1,3)$ ribbon contained 120 / 66 atoms corresponding to a length of 25.75 A / 12.86 A, respectively. The simulation cell contained 15 Å of vacuum in the non-periodic directions to minimize interactions among periodic replica of the system. The thickness of the vacuum region, the sampling of the BZ, and the cutoff ensure convergence for the computed band structures. The atomic positions of the ribbon and the cell dimension along the ribbon axis were optimized until forces were lower than 0.002 eV/Å and the pressure in the cell was negligible. The DFT band structures shown in Fig. S12 are aligned to the vacuum level, which was computed from the average electrostatic potential in the vacuum region. All calculations were performed through a Jupiter based GUI for the AiiDA platform [8].

## 3. Magnetic properties of the edge-extended AGNRs

The zigzag edge structure at the boundary of the N-AGNR backbone and the edge-extended ribbon segments suggests the possibility of magnetic ordering of the corresponding boundary states. To explore this, Mean-Field Hubbard Tight Binding (MFH-TB) calculations were carried out in the single-band scheme by considering only the nearest-neighbor hopping parameter $\gamma_0$ = 3 eV in the tight binding term of the Hamiltonian, and an on-site Coulomb repulsion $U = 1 * \gamma_0$ [9,10]. Half filling, i.e. charge neutrality, was assumed for all calculations. The iterative self-consistent solution of the Hubbard Hamiltonian was initiated with a weak antiferromagnetic (AFM) bias on the zigzag edges of 0.05 µB per atom in all cases.

The localized nature of the spin-degenerate zero energy junction states implies that they can be susceptible to Hubbard type splitting due to Coulombic electron-electron interaction. This energy splitting, which is the analogue of the opening of the Hubbard gap in regular zigzag edge GNRs, is accompanied with spin polarization of the corresponding states [9]. The effect is also analogous to the spin polarization occurring at the short zigzag ends of $N$-AGNRs for $N \geq 7$ [11].

a                                                                        b

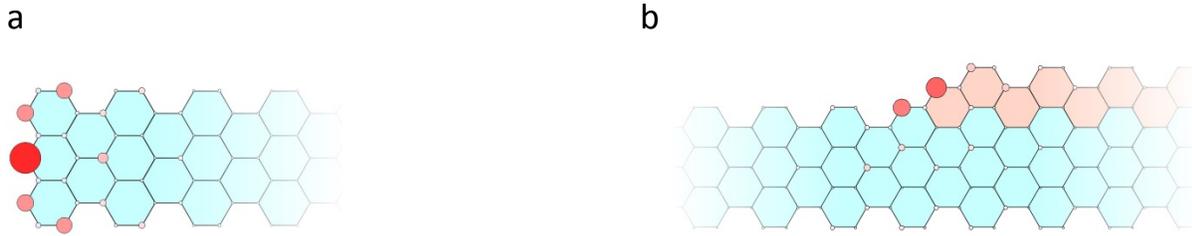

**Figure S13**
Spin polarization at **a**, the short zigzag end of a 7-AGNR, and **b**, at the 7-ANGR / 9-AGNR junction. Marker size denotes the absolute value of the spin polarization $|n_{i,\uparrow} - n_{i,\downarrow}|$ and marker color the relative value of the spin imbalance $\mu_i = n_{i,\uparrow} - n_{i,\downarrow}$ (red -1/3 µB to blue +1/3 µB).

For the two examples shown in Fig. S13 the maximum spin polarization occurs at the center of the 3 atom short zigzag edges and amounts to $|\mu_{max}|$=0.28 µB and $|\mu_{max}|$=0.20 µB for the pristine 7-AGNR end state (Fig. S13a**)** and the 7-AGNR – 9-AGNR boundary state (Fig. S13b), respectively. The total absolute spin anisotropy $\mu_{tot} = \sum |n_{i,\uparrow} - n_{i,\downarrow}|$ is µtot=1.64 µB and µtot=1.73 µB for the two cases. Here $n_{i,\uparrow}$ denotes the spin-up density on the atomic site $i$ of the structure. The Hubbard energy splitting of the up- and down-spin localized state is 0.129 $\gamma_0$ and 0.070 $\gamma_0$, respectively.
This Hubbard type spin dependent energy splitting and the associated spin polarization of the boundary state is strongly suppressed upon hybridization with neighboring boundary states or metallic states. One should thus only expect sizeable spin polarization and energy splitting for $N = 3p + 1$ and large $n$ and $m$.

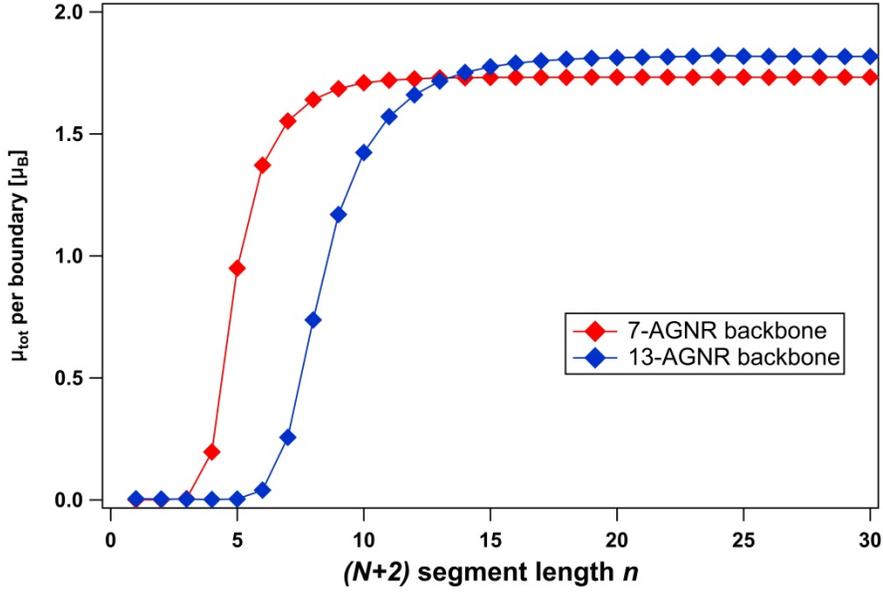

**Figure S14**

Total absolute magnetization $\mu_{tot}$ per $N$-AGNR / $(N+2)$-AGNR boundary state as a function of segment length $n$ with $m \sim \infty$ for a 7-AGNR (red) and a 13-AGNR backbone (blue).

Fig. S14 reveals that the interaction of the left and right boundary state for short (i.e. $n < 4$) edge-extended $(N+2)$-AGNR segments is so strong that the Hubbard splitting is effectively suppressed, as evidenced by suppression of the spin polarization (for an on-site Coulomb repulsion chosen as $U_C = 1 * \gamma_0 = 3$ eV). This means that the experimentally realized structures ($n$=1) are essentially non-magnetic, also if values of $U_C$ up to $2\gamma_0$ are considered. For structures of larger $n$ and $m$ (i.e. smaller $t_n$ and $t_m$), however, the system will exhibit spin polarization. Figure S15 shows the spin polarization of a 7-AGNR-$S(2,4)$ revealing formation of a staggered, antiferromagnetic spin chain for a high, yet not unrealistic [12] value of $U_C = 1.6 * \gamma_0$.

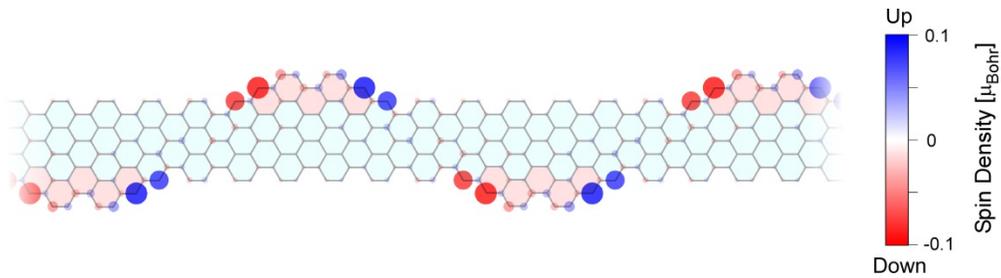

**Figure S15**

Spin polarization expressed by the spin-density $(n_\uparrow - n_\downarrow)$ for the 7-AGNR-$S(2,4)$ with $U_C = 1.6 * \gamma_0$. Marker color and size denote the magnitude $(n_\uparrow - n_\downarrow)$ and absolute magnitude $(|n_\uparrow - n_\downarrow|)$ of the spin polarization.

## 4. Conservation of the topological end states in finite $S$ and $I$ edge-extended AGNR structures

The terminus of a $N$-AGNR-$S(n,m)$ can host two kinds of zero energy states: A topological end state originating from the segment boundary states, and the zigzag end state of the N-AGNR backbone (Fig. S13a) [11]. Therefore, the E=0 states occurring at the ends of non-trivial $N$-AGNR-$S(n,m)$ and $N$-AGNR-$I(n,m)$ structures need to be protected from hybridization with the zigzag end states of the backbone $N$-AGNR. Figure S16 shows the end of a 7-AGNR-$S(1,1)$ structure (a topologically non-trivial insulator) that is longitudinally extended by 8 units (Fig. S16a) or 2 units (Fig. S16b) of the backbone 7-AGNR.

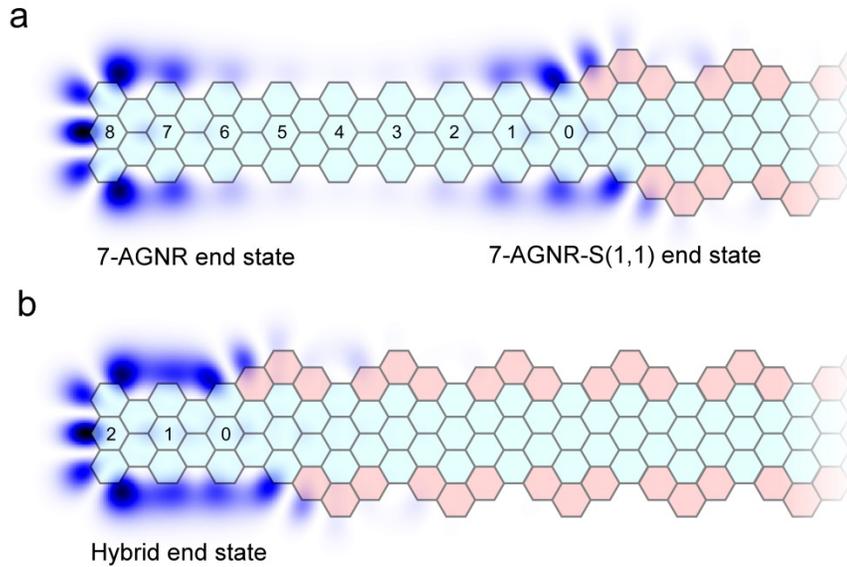

**Figure S16**
**a**, 7-AGNR- $S(1,1)$ structure with terminus longitudinally extended by 8 units of the 7-AGNR backbone. The charge density of the frontier orbitals close to E=0 are overlaid to the structural drawing. **b**, Same as (a) but for a shorter end extension of only 2 backbone units.

Fig. S16a reveals that the zigzag end state of the 7-AGNR backbone and the topological end state of the 7-AGNR-$S(1,1)$ show only marginal (if any) orbital overlap for reasonably long backbone extension (8 units). The situation is different if the backbone extension length is reduced to 2, as shown in Fig. S16b. In this case the two end states overlap strongly and split in energy. For the inline $N$-AGNR-$I(n,m)$ structures the situation is analogous.

The dependence on backbone extension length of the hybridization between the backbone end state and the end state of topologically non-trivial staggered and inline AGNRs is illustrated in Fig. S17 for a 7-AGNR-$S(1,1)$ and a 7-AGNR-$I(1,3)$. For both hybrid structures it can be observed, that a backbone extension exceeding 7 or 8 7-AGNR units effectively suppresses hybridization of the two different end states. For backbone extensions of less than 5 7-AGNR units, however, a sizeable bonding and anti-bonding splitting of the zero-energy end states is observed. In the case of the experimentally realized 7-AGNR-$I(1,3)$ structure the end states are completely suppressed for backbone extensions shorter than 2 units.

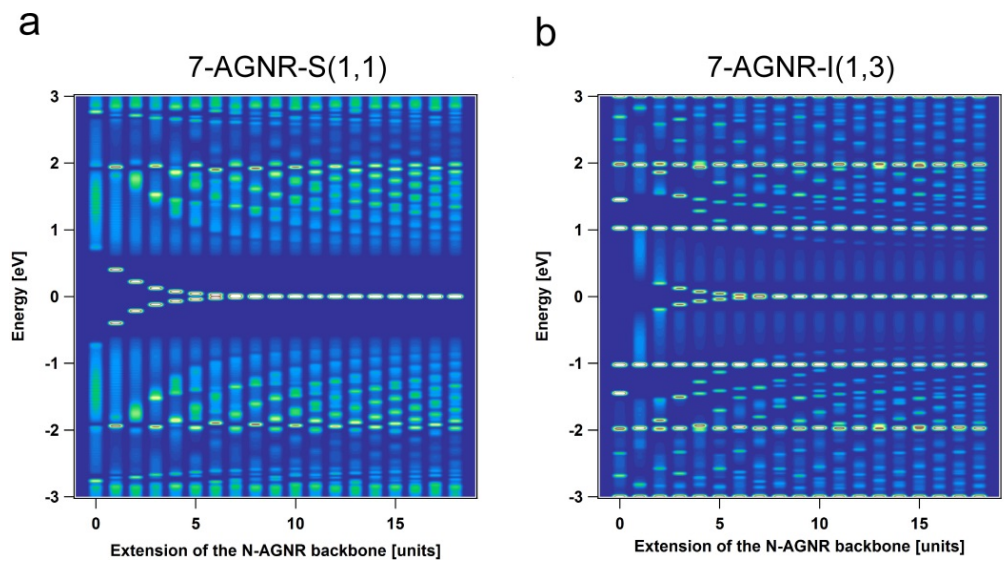

**Figure S17**

LDOS at the junction end state as a function of the length of the backbone extension in (terms of 7-AGNR unit cells) for **a**, the 7-AGNR-$S(1,1)$ and **b**, the 7-AGNR-$I(1,3)$.

In the case of a topologically trivial structure such as the experimentally realized 7-AGNR-$S(1,3)$, only the zigzag backbone end state is present. Due to particle-hole symmetry, it remains at E=0 independent of the length of the backbone extension. This can clearly be seen in Fig. S18a where the LDOS at the short zigzag 7-AGNR terminus is plotted as a function of the backbone extension length. Even in the absence of any backbone extension the zero-energy state persists and can be observed experimentally, as is confirmed in Fig. S18d.

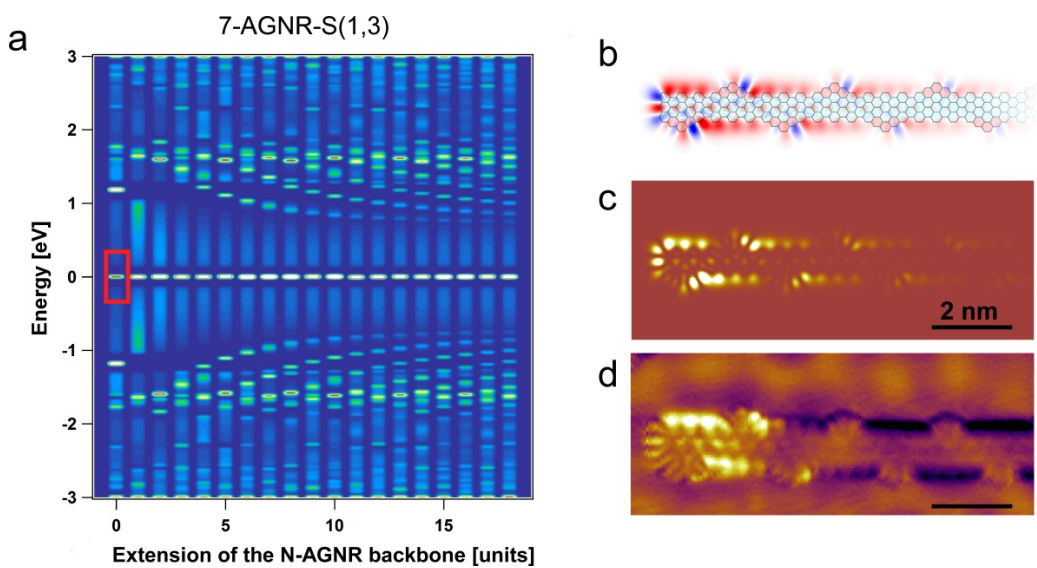

**Figure S18**

**a**, LDOS map of the 7-AGNR-$S(1,3)$ structure at the zigzag terminus as a function of 7-AGNR backbone extension length. **b**, Wave function map of the E=0 state for a 7-AGNR-$S(1,3)$ (without backbone extension), and **c**, corresponding TB-STS simulation of the E=0 state (marked by red rectangle in a)). **d**,

Experimental dI/dV map acquired from a corresponding 7-AGNR-$S(1,3)$ end segment at +0.15 eV (500 pA).

## 5. On-surface synthesis

Synthesis of the 7-AGNR-$S(1,3)$ and 7-AGNR-$I(1,3)$ nanoribbon structures has been realized by means of dehalogenative aryl-aryl coupling and subsequent cyclodehydrogenation of the precursor monomers **1** and **3** (for details see section 7. "Precursor synthesis"), respectively, on a Au(111) surface under ultrahigh-vacuum conditions. The on-surface synthesis strategy is analogous to the one described by Cai et al. [13] for 7-AGNRs, but complemented by the use of methyl groups that undergo cyclization with the neighboring aromatic rings to form smooth zigzag edges.

Au(111) single crystal substrates were cleaned by standard argon ion sputtering and annealing cycles. Molecular precursors were sublimated using a homemade six-fold evaporator, which allows sublimation of two molecules sequentially or at the same time. The deposition flux was around 2Å/min as determined by a quartz crystal microbalance.

7-AGNR-$S(1,3)$ nanoribbons were synthesized via a multiple step annealing process to achieve high-quality samples. The molecular precursor (monomer **1**, 6,11-bis(10-bromoanthracene-9-yl)-1,4-dimethyltetracene, BADMT) was deposited on a clean Au(111) substrate held at room temperature. Subsequently, the sample was annealed at 200° C, 250° C, 280° C and 400° C for 10 minutes at each temperature. This procedure results in the stepwise monomer activation (i.e. radical formation by dehalogenation), polymerization of the radical intermediates, and subsequent cylcodehydrogenation to form the fully conjugated AGNR structure. GNRs of lengths up to 60 nm with excellent quality could be grown on Au(111), as is illustrated in Fig S19.

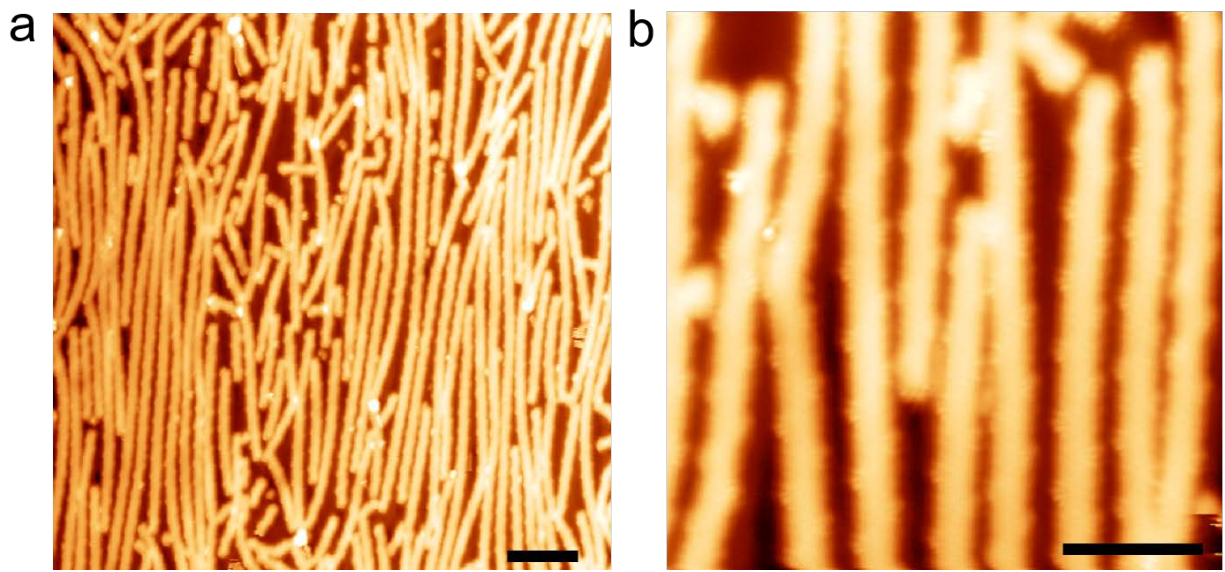

**Figure S19**
**a**, Large scale constant current STM image of 7-AGNR-$S(1,3)$ ribbons grown on Au(111). Scale bar: 10 nm. **b**, Zoomed-in constant current STM image of 7-AGNR-$S(1,3)$ ribbons. Scale bar: 5 nm.

Fig. S19 and Fig. S20 reveal that the alternating sequence of the dimethyltetracene units which leads to the staggered 7-AGNR-$S(1,3)$ structure is perfect. This indicates a very high conformational selectivity in the precursor coupling process, details of which will be discussed elsewhere [14].

On-surface synthesis of backbone extended 7-AGNR / 7-AGNR-$S(1,3)$ heterostructures was achieved via sequential deposition of monomer **1** and monomer **2** (10,10'-dibromo-9,9'-bianthryl, DBBA). In a first step, monomer **1** was deposited on the Au(111) substrate held at room temperature, followed by an annealing step at 150° C for 10 minutes. In a second step, monomer **2** was deposited onto the sample held at 150° C, followed by annealing steps at 200° C, 250° C, and 400° C for 10 minutes at each temperature. With this recipe, the 7-AGNR segments always locate at the termini of heterostructures, as seen in Fig. S20.

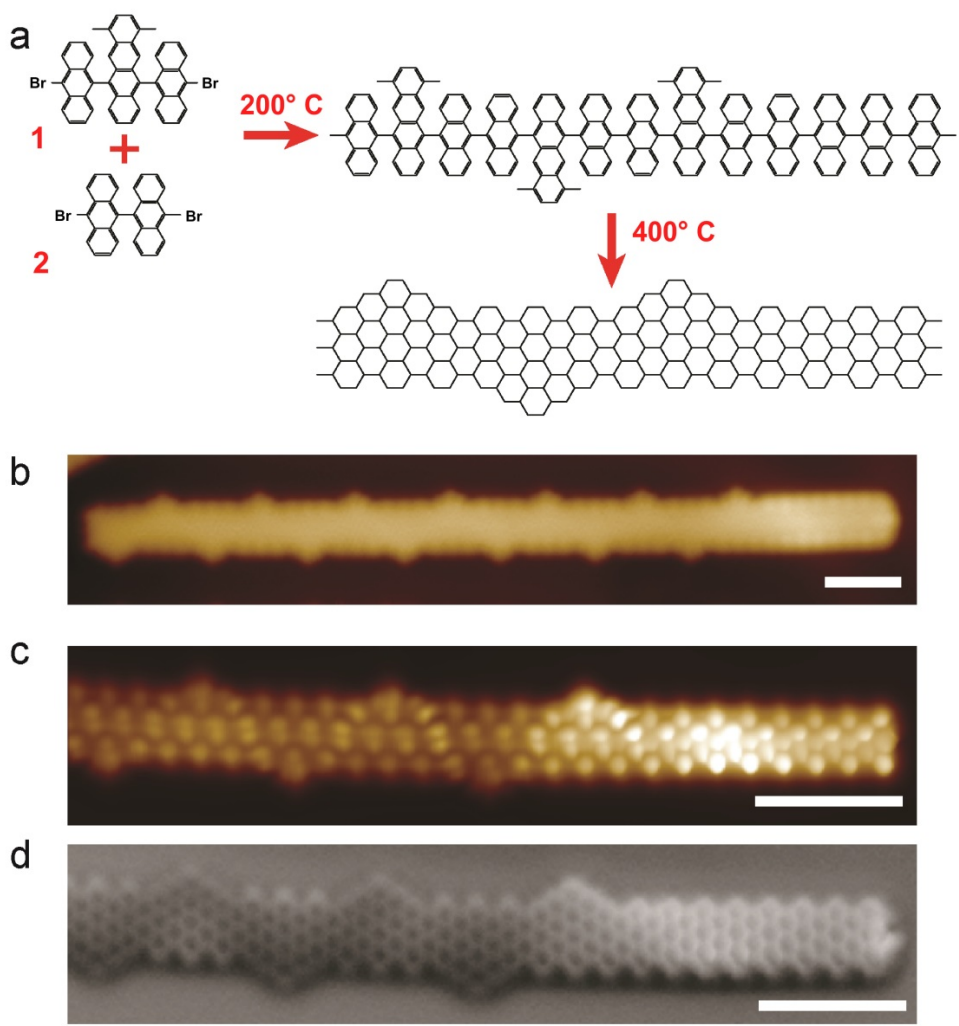

**Figure S20**
**a**, Schematic representation of the synthetic pathway to the backbone extended 7-AGNR – 7-AGNR-$S(1,3)$ heterostructure. **b**, Constant-current STM image (-1.5 V / 10 pA) of a single 7-AGNR-$S(1,3)$ ribbon whose right terminus is extended by a short 7-AGNR backbone segment. **c** and **d**, constant-height nc-AFM images of the tunneling current at 10 mV (c) and the frequency shift (d) of the hybrid structure in **b**. The scale bar is 2 nm in all panels.

7-AGNR / 7-AGNR-$I(1,3)$ heterostructures were synthesized using the same recipe than the one described above for 7-AGNR / 7-AGNR-$S(1,3)$ heterostructures, but with the precursor monomers **2** and **3** (6,13-bis(10-bromoanthracene-9-yl)-1,4,8,11-tetramethylpentacene, BATMP) (Fig. 21a). High resolution STM and nc-AFM images of a resulting heterostructure are shown in Fig. S21b and S21c, respectively. A high resolution STS dI/dV map acquired at +0.15 V clearly reveals the topological end state at the terminus of the 7-AGNR-$I(1,3)$ (Fig. S21d). The corresponding TB simulation of the end state charge density (Fig. S21e) and the corresponding end state wave function (Fig. S21f) show excellent agreement with experimental data.

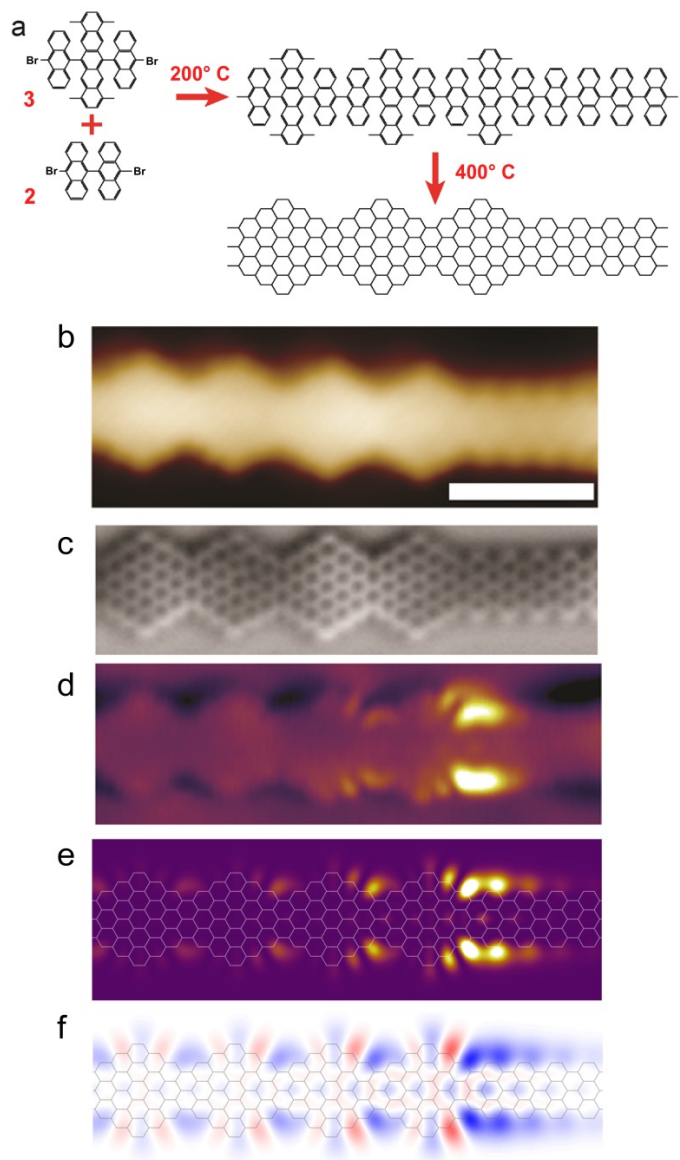

**Figure S21**
**a**, Schematic representation of the synthetic pathway to the backbone extended 7-AGNR – 7-AGNR-$I(1,3)$ heterostructure. **b**, Constant-current STM image (10 mV / 50 pA) of a single 7-AGNR-$I(1,3)$ ribbon whose right terminus is extended by a 7-AGNR backbone segment. **c**, Constant-height frequency shift nc-AFM image of the heterostructure shown in (a). **d**, Constant-current STS dI/dV map (at +0.25 V / 500 pA) of the topological end state. **e**, TB computed charge density map of the topological end state superimposed on a structural model. **f**, Wave function map associated with (e). The scale bar is 2 nm.

## 6. Raman characterization

Experimental Raman data from a sample of 7-AGNR-$S(1,3)$ grown on Au(111) were obtained under ambient conditions using a Raman Bruker Senterra spectrometer (3 × 60 seconds, 2 mW) with an incident wavelength of 532 nm.

Raman modeling of the 7-AGNR-$S(1,3)$ was conducted with a combination of density functional theory (DFT) based calculations to determine the normal modes of the structure and of a bond-polarization model for the determination of Raman intensities. The DFT calculations were performed with the VASP package [15] using projector-augmented-wave (PAW) pseudopotentials, the Perdew–Burke–Ernzerhof (PBE) exchange-correlation functional [6] and a plane-wave cutoff energy of 400 eV. Prior to building the dynamical matrix for the calculation of phonons, all the atomic positions were relaxed until residual forces were below 0.001 eV/Å. Phonon calculations were performed utilizing the Phonopy package, written by Atsushi Togo [16]. From the atomic positions and phonon normal mode information, non-resonant Raman spectrum intensities were calculated using an empirical bond-polarization model [17].

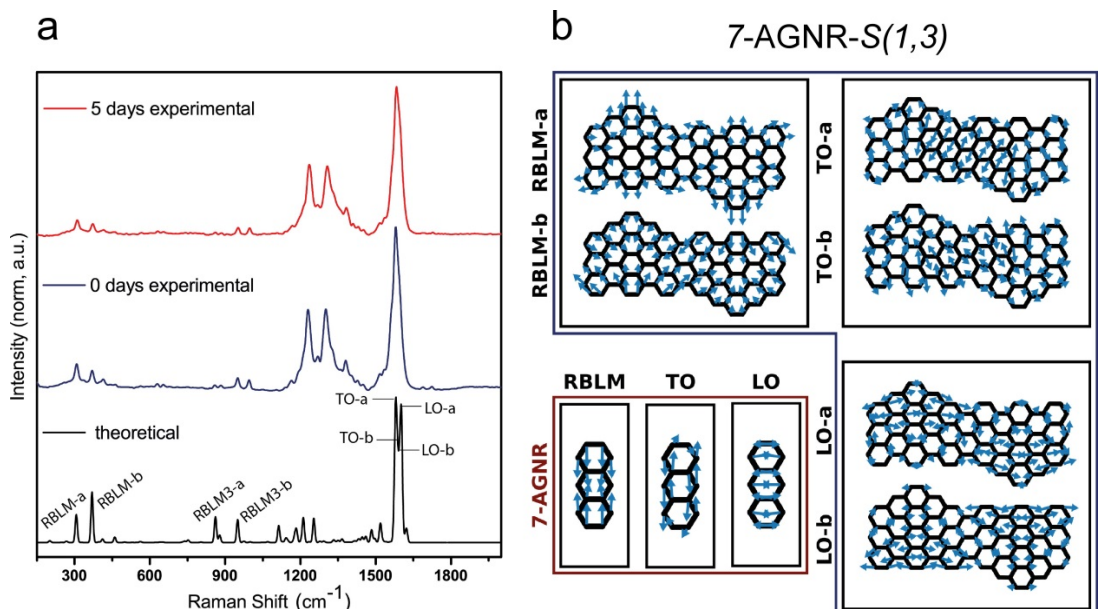

**Figure S22**
Raman analysis of the staggered edge-extended 7-AGNR-$S(1,3)$ ribbon structure. **a**, Comparison of DFT simulated spectrum (black curve) and experimental Raman spectra taken directly after growth (blue curve) and after 5 days in air (red curve). Both experimental and simulated spectra show multiple radial breathing like modes (RBLM) centered at about 300 cm$^{-1}$ (experiment: 308 and 370 cm$^{-1}$; DFT: 307 and 370 cm$^{-1}$, labeled RBLM-a and RBLM-b) and a multitude of TO and LO mode analogues. These are located around 1580 cm$^{-1}$ (experiment) and at 1579, 1587, 1595, and 1603 cm$^{-1}$ (DFT) and labeled TO-a, TO-b, LO-a, and LO-b, respectively. These rich features are unique fingerprints of GNRs with non-trivial edge structure. The good agreement between experimental and theoretical data underlines the quality of the sample. A 5 days stability study proves the robustness of the 7-AGNR-$S(1,3)$ under ambient conditions, with no significant changes in the main Raman modes ("5 days experimental", measured under the same conditions as "0 days experimental"). **b**, Phonon mode schematics of selected Raman-active modes for the staggered edge-extended 7-AGNR-$S(1,3)$ as well as for the pristine 7-AGNR for comparison.

## 7. Precursor synthesis

**General**

Column chromatography was conducted with silica gel (grain size 0.063–0.200 mm or 0.04–0.063 mm) and thin layer chromatography (TLC) was performed on silica gel-coated aluminum sheets with F254 indicator. Nuclear Magnetic Resonance (NMR) spectra were recorded on a Bruker Avance 300 MHz spectrometer. Chemical shifts were reported in ppm. Coupling constants (*J* values) were presented in Hertz. $^1$H NMR chemical shifts were referenced to $C_2D_2Cl_4$ (6.00 ppm). $^{13}$C NMR chemical shifts were referenced to $C_2D_2Cl_4$ (73.78 ppm). Abbreviations: s = singlet, d = doublet, dd = double doublet, t = triplet, m = multiplet. High-resolution mass spectrometry (HRMS) was performed on a SYNAPT G2 Si high resolution time-of-flight mass spectrometer (Waters Corp., Manchester, UK) by matrix-assisted laser desorption/ionization (MALDI). Melting points were measured with a Büchi B-545 apparatus.

**Materials**

All commercially available chemicals were purchased from TCI, Aldrich, Acros, Merck, and other commercial suppliers and used without further purification unless otherwise noted. Monomer **2** (10,10'-dibromo-9,9'-bianthryl) was synthesized according to our previous report [13]. 2,3-Bis(bromomethyl)-1,4-dimethylbenzene (**4**) and 1,4,8,11-tetramethyl-6,13-pentacenedione (**6**) were prepared following reported procedures [18,19].

**Synthesis of monomers 1 and 3**

Monomers **1** and **3** were synthesized as shown in Scheme S1. 7,10-Dimethyltetracene-5,12-dione (**5**) was initially prepared through Diels–Alder cycloaddition of 1,4-napthoquinone and an *o*-quinodimethane *in-situ* generated from 2,3-bis(bromomethyl)-1,4-dimethylbenzene (**4**) through iodide-induced debromination [18]. Subsequently, **5** was reacted with 10-bromo-9-anthracenyllithium generated by mono-lithiation of 9,10-dibromoanthracene, followed by dehydroxylation with sodium iodide/sodium hypophosphite monohydrate to give monomer **1** in 40% yield over two steps. Synthesis of monomer **3** was conducted in a similar manner, through the reaction of 1,4,8,11-tetramethyl-6,13-pentacenedione (**6**) [19] with 10-bromo-9-anthracenyllithium, followed by the dehydroxylation to afford monomer **3** in 18% yield over two steps.

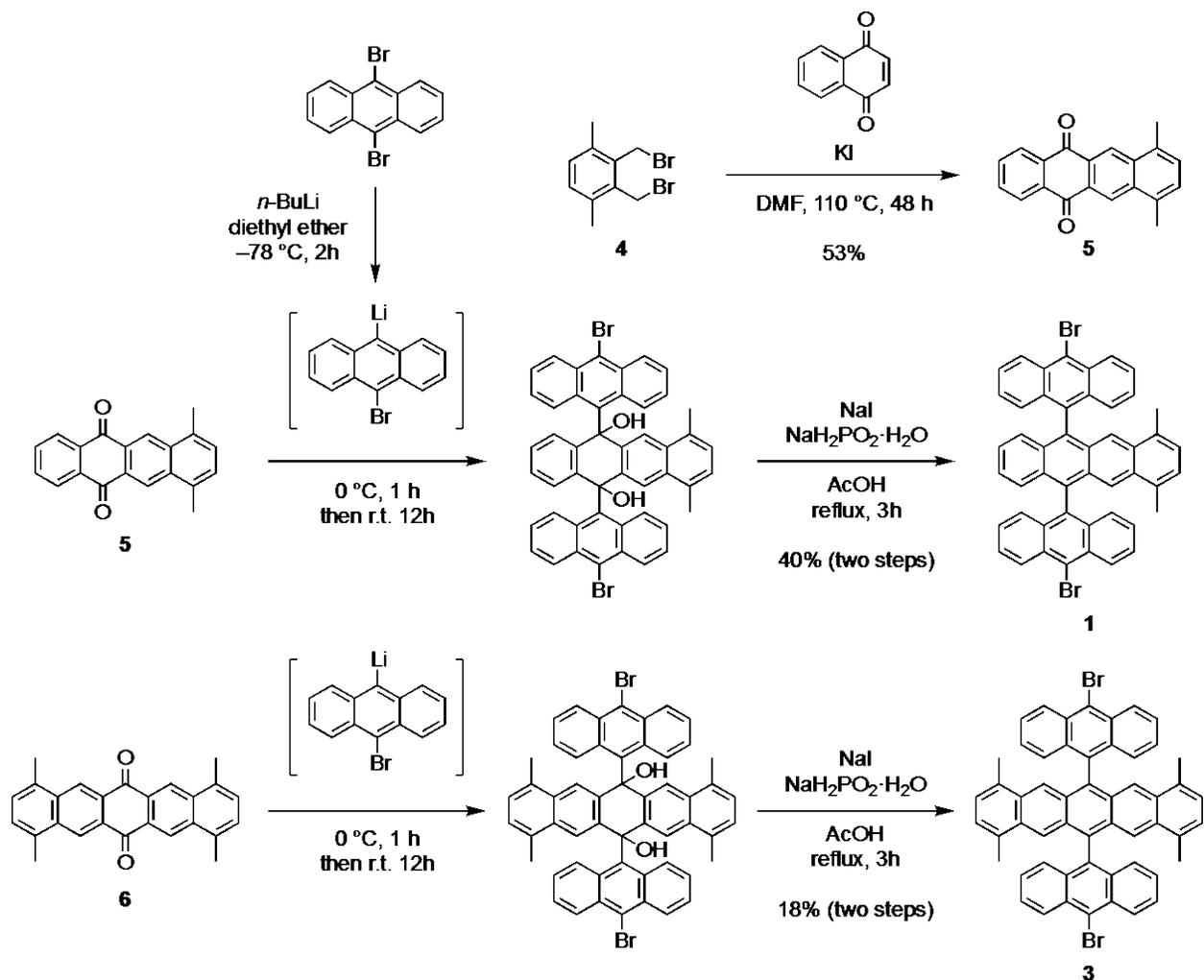

**Scheme S1**. Synthesis of monomers **1** and **3**.

**7,10-dimethyltetracene-5,12-dione (5)**

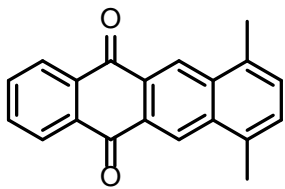

A mixture of 2,3-bis(bromomethyl)-1,4-dimethylbenzene (**4**) (3.00 g, 10.3 mmol), 1,4-naphthoquinone (1.62 g, 10.3 mmol) and KI (17.1 g, 103 mmol) in dry dimethylformamide (DMF) (50 mL) was stirred at 110 °C for 48 h. The reaction mixture was cooled to room temperature and then 100 mL of water was added. After filtration, the resulting precipitates were intensively washed with water, ethanol, ethyl acetate and hexane to obtain the title compound as an orange-yellow solid (1.54 g, 53% yield). M.p.: 273.4–274.6 °C. $^1$H NMR (300 MHz, $C_2D_2Cl_4$, 298 K, ppm) $\delta$ 8.89 (d, $J$ = 1.3 Hz, 2H), 8.32 (ddd, $J$ = 5.9, 3.3, 1.3 Hz, 2H), 7.81 (ddd, $J$ = 5.9, 3.3, 1.3 Hz, 2H), 7.36 (d, $J$ = 1.3 Hz, 2H), 2.75 (d, $J$ = 1.3 Hz, 6H); $^{13}$C NMR (75 MHz, $C_2D_2Cl_4$, 298 K, ppm) $\delta$ 183.09, 135.23, 134.73, 134.35, 134.27, 130.22, 128.70, 127.37, 126.28, 19.48. HRMS (MALDI-TOF, positive) *m/z*: Calcd for $C_{20}H_{14}O_2$: 286.0994; Found: 286.0989 [M]$^+$.

**6,11-bis(10-bromoanthracen-9-yl)-1,4-dimethyltetracene (1)**

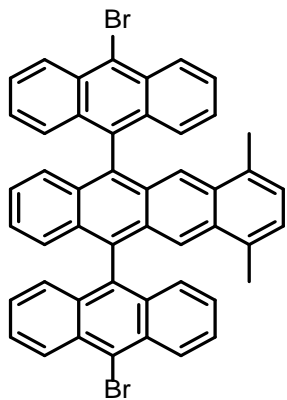

To a suspension of 9,10-dibromoanthracene (1.01 g, 3.00 mmol) in dry diethyl ether (30 mL) was added a solution of *n*-Butyllithium (*n*-BuLi) (1.5 mL, 2.4 mmol, 1.6 M in hexanes) dropwise at −78 °C. The reaction mixture was stirred for 2 h under argon to obtain an orange solution of 10-bromo-9-anthracenyllithium. To a suspension of 7,10-dimethyltetracene-5,12-dione (**7**) (258 mg, 0.900 mmol) in 20 mL of dry diethyl ether was added at 0 °C the separately prepared solution of 10-bromo-9-anthracenyllithium via a double-tipped needle under argon. The reaction mixture was stirred for 1 h at this temperature, and then allowed to gradually warm to room temperature. After stirring for 12 h, the reaction was quenched by adding 3 mL of glacial acetic acid. The precipitates were then collected by filtration and washed with diethyl ether. To the resulting crude material placed in a 100-mL round-bottom flask was added glacial acetic acid (45 mL), NaI (1.35 g, 9.00 mmol) and $NaH_2PO_2 \cdot H_2O$ (1.43 g, 13.5 mmol). The reaction mixture was refluxed for 3 h under the exclusion of light. After cooling down to room temperature, the precipitates were collected by filtration, and then washed with water and methanol to afford analytically pure monomer **1** as an orange solid (288 mg, 40% yield). For the on-surface experiments, further purification was performed by recrystallization through slow diffusion of degassed methanol into a degassed solution of **1** in dichloromethane under argon, which was repeated 8 times. M.p.: >315 °C. $^1$H NMR (300 MHz, $C_2D_2Cl_4$, 298 K, ppm) δ 8.80 (d, *J* = 9.1 Hz, 4H), 7.99 (s, 2H), 7.71 – 7.67 (m, 4H), 7.39 (d, *J* = 8.8 Hz, 4H), 7.33 (dt, *J* = 9.9, 4.6 Hz, 4H), 7.22 (d, *J* = 8.1 Hz, 2H), 7.09 (d, *J* = 8.2 Hz, 2H), 6.91 (s, 2H), 2.01 (s, 6H); $^{13}$C NMR (75 MHz, $C_2D_2Cl_4$, 298 K, ppm) δ 132.48, 130.53, 128.14, 127.46, 127.16, 126.55, 125.85, 125.66, 123.18, 19.15. HRMS (MALDI-TOF, positive) *m/z*: Calcd for $C_{48}H_{30}Br_2$: 764.0714; Found: 764.0705 [M]$^+$.

**6,13-bis(10-bromoanthracen-9-yl)-1,4,8,11-tetramethylpentacene (3)**

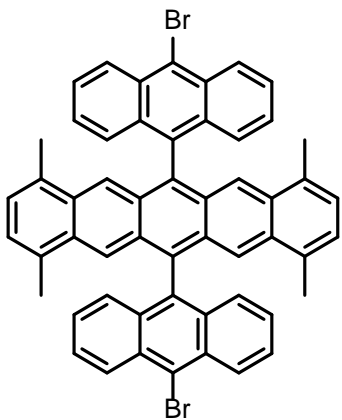

To a suspension of 9,10-dibromoanthracene (300 mg, 0.893 mmol) in dry diethyl ether (20 mL) was added a solution of *n*-BuLi (0.45 ml, 0.72 mmol, 1.6 M in hexanes) dropwise at –78 °C. The reaction mixture was stirred for 2 h under argon to obtain an orange solution of 10-bromo-9-anthracenyllithium. To a suspension of 1,4,8,11-tetramethyl-6,13-pentacenedione (**6**) (97.6 mg, 0.268 mmol) in 10 mL of dry diethyl ether was added at 0 °C the separately prepared solution of 10-bromo-9-anthracenyllithium via a double-tipped needle under argon. The reaction mixture was stirred for 1 h at this temperature, and then allowed to gradually warm to room temperature. After stirring for 12 h, the reaction was quenched by adding 5 mL of glacial acetic acid. The precipitates were then collected by filtration and washed with diethyl ether. To the resulting crude material placed in a 100-mL round-bottom flask was added glacial acetic acid (30 mL), NaI (405 mg. 2.70 mmol) and $NaH_2PO_2 \cdot H_2O$ (429 mg. 4.05 mmol). The reaction mixture was refluxed for 3 h under the exclusion of light. After cooling down to room temperature, the precipitates were collected by filtration, and then washed with water and methanol to afford analytically pure monomer **3** as a dark blue solid (140 mg, 18% yield). For the on-surface experiments, further purification was performed by recrystallization through slow diffusion of degassed methanol into a degassed solution of **3** in chloroform under argon, which was repeated 7 times. M.p.: >315 °C. $^1$H NMR (300 MHz, $C_2D_2Cl_4$, 298 K, ppm) $\delta$ 8.82 (d, *J* = 8.9 Hz, 2H), 7.98 (s, 2H), 7.68 (t, *J* = 7.8 Hz, 2H), 7.38 (d, *J* = 8.8 Hz, 2H), 7.29 (t, *J* = 7.7 Hz, 2H), 6.85 (s, 2H), 1.99 (s, 6H); $^{13}$C NMR (75 MHz, $C_2D_2Cl_4$, 298 K, ppm) $\delta$ 132.50, 130.52, 128.09, 127.65, 127.40, 126.51, 125.56, 123.04, 19.12. HRMS (MALDI-TOF, positive) *m/z*: Calcd for $C_{54}H_{36}Br_2$: 842.1184; Found: 842.1197 $[M]^+$.

**Single crystal X-ray diffraction analysis**

The single crystal of **1** suitable for X-ray analysis was obtained by slow diffusion of degassed methanol into a degassed solution of **1** in dichloromethane under argon and exclusion of light.

*Crystal data*

| | |
|---|---|
| formula | $C_{48}H_{30}Br_2$ |
| molecular weight | 766.54 gmol$^{-1}$ |
| absorption | $\mu$ = 2.049 mm$^{-1}$ correction with 7 crystal faces |
| transmission | Tmin = 0.7275, Tmax = 0.9053 |
| crystal size | 0.06 x 0.06 x 0.24 mm$^3$ brown needle |
| space group | P 2$_1$/n (monoclinic) |
| lattice parameters | a = 9.0127(6)Å |
| (calculate from | b = 23.8476(11)Å    ß = 93.029(6)° |
| 14422 reflections with | c = 18.7144(13)Å |

| | | |
|---|---|---|
| 2.2° < θ < 28.1°) | $V = 4016.7(4) Å^3$  $z = 4$ | $F(000) = 1552$ |
| temperature | -80°C | |
| density | $d_{xray} = 1.268$ gcm$^{-3}$ | |

*Data collection*

| | |
|---|---|
| diffractometer | STOE IPDS 2T |
| radiation | Mo-K$_α$ Graphitmonochromator |
| Scan – type | ω scans |
| Scan – width | 1° |
| scan range | 2° ≤ θ < 28° |
| | $-11 ≤ h ≤ 9$  $-31 ≤ k ≤ 27$  $-24 ≤ l ≤ 24$ |
| number of reflections: | |
| measured | 22569 |
| unique | 9881 ($R_{int} = 0.0688$) |
| observed | 4323 ($|F|/σ(F) > 4.0$) |

*data correction, structure solution and refinement*

| | |
|---|---|
| corrections | Lorentz and polarisation correction |
| Structure solution | Program: SIR-2004 (Direct methods) |
| refinement | Program: SHELXL-2014 (full matrix). 454 refined parameters, weighting scheme: $w=1/[σ^2(F_o^2) + (0.1088*P)^2]$ with $(Max(F_o^2,0)+2*F_c^2)/3$. H-atoms at calculated positions and refined with isotropic displacement parameters, non H- atoms refined anisotropically. |
| R-values | wR2 = 0.22  (R1 = 0.0687 for observed reflections, 0.1593 for all reflections) |
| goodness of fit | S = 0.943 |
| maximum deviation of parameters | 0.001 * e.s.d |
| maximum peak height in diff. Fourier synthesis | 0.65, -0.89  eÅ$^{-3}$ |
| remark | structure contains two molecules CH$_2$Cl$_2$ which are completely disordered – SQUEEZE was used |

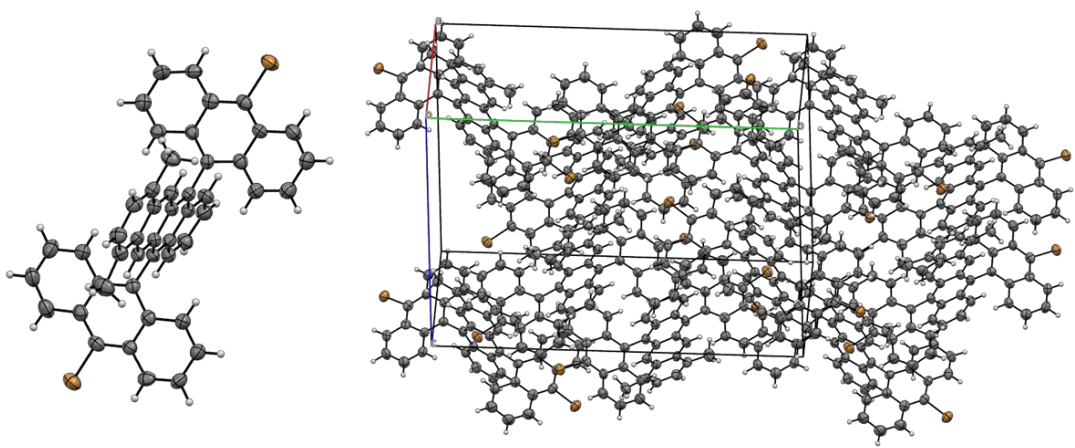

**Figure S23**
X-ray single-crystal analysis of **1** (ORTEP drawings with thermal ellipsoids set at 50% probability). Bromine atoms are labelled in brown.

The single crystal of **3** suitable for X-ray analysis was obtained by slow diffusion of degassed methanol into a degassed solution of **3** in chloroform under argon and exclusion of light .

*Crystal data*

| | |
|---|---|
| formula | $C_{54}H_{36}Br_2$, $1(CHCl_3)$ |
| molecular weight | 1083.38 gmol$^{-1}$ |
| absorption | $\mu$ = 2.129 mm$^{-1}$ correction with 6 crystal faces |
| transmission | $T_{min}$ = 0.480, $T_{max}$ = 0.934 |
| crystal size | 0.04 x 0.05 x 0.64 mm$^3$ brown needle |
| space group | P -1 (triclinic) |
| lattice parameters | a = 8.4370(6)Å    $\alpha$ = 113.069(5)° |
| (calculate from | b = 12.4974(8)Å   ß = 103.666(5)° |
| 7729 reflections with | c = 13.0211(8)Å   $\gamma$ = 101.311(6)° |
| 2.6° < θ < 28.4°) | V = 1161.63(14)Å$^3$  z = 1  F(000) = 546 |
| temperature | -80°C |
| density | $d_{xray}$ = 1.549 gcm$^{-3}$ |

*Data collection*

| | |
|---|---|
| diffractometer | STOE IPDS 2T |
| radiation | Mo-K$_\alpha$ Graphitmonochromator |
| Scan – type | ω scans |
| Scan – width | 1° |
| scan range | 2° ≤ θ < 28° |
| | -11 ≤ h ≤ 10   -16 ≤ k ≤ 16   -17 ≤ l ≤ 17 |
| number of reflections: | |
| measured | 11053 |
| unique | 5729 ($R_{int}$ = 0.0448) |
| observed | 3711 (|F|/σ(F) > 4.0) |

*data correction, structure solution and refinement*

| | |
|---|---|
| corrections | Lorentz and polarisation correction |
| Structure solution | Program: SIR-2004 (Direct methods) |
| refinement | Program: SHELXL-2014 (full matrix). 291 refined parameters, weighting scheme: $w=1/[\sigma^2(F_o^2) + (0.0477*P)^2+3.26*P]$ with $(Max(F_o^2,0)+2*F_c^2)/3$. H-atoms at calculated positions and refined with isotropic displacement parameters, non H- atoms refined anisotropically. |
| R-values | wR2 = 0.1483  (R1 = 0.0576 for observed reflections, 0.1064 for all reflections) |
| goodness of fit | S = 1.034 |
| maximum deviation of parameters | 0.001 * e.s.d |
| maximum peak height in diff. Fourier synthesis | 0.51, -0.55 eÅ$^{-3}$ |
| remark | crystal contains two molecules of solvents per main molecule, which has Ci symmetry |

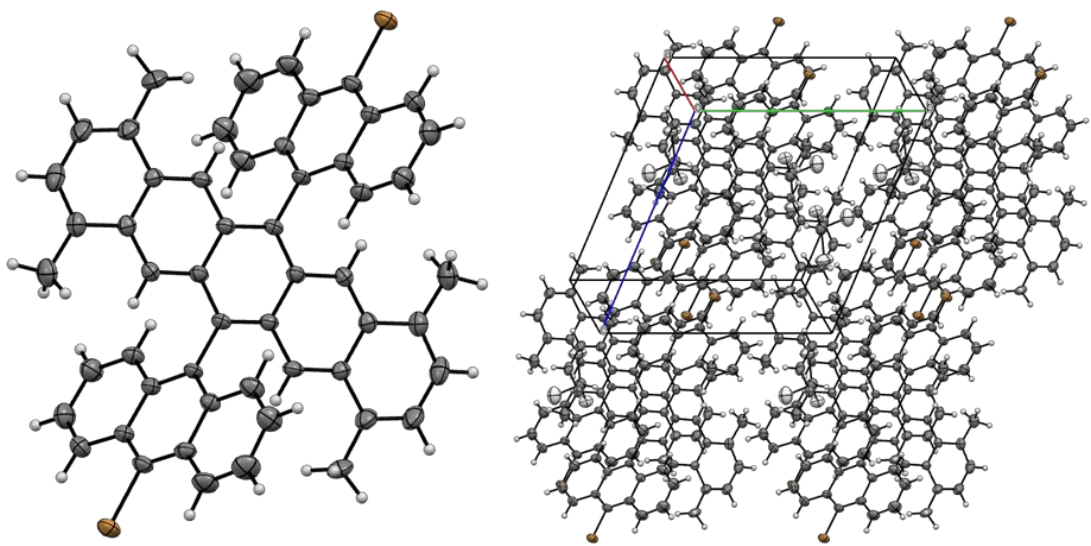

**Figure S24**

X-ray single-crystal analysis of **3** (ORTEP drawings with thermal ellipsoids set at 50% probability). Solvent molecules (CHCl$_3$) are omitted for clarity. Bromine atoms are labelled in brown.

**NMR and mass spectra**

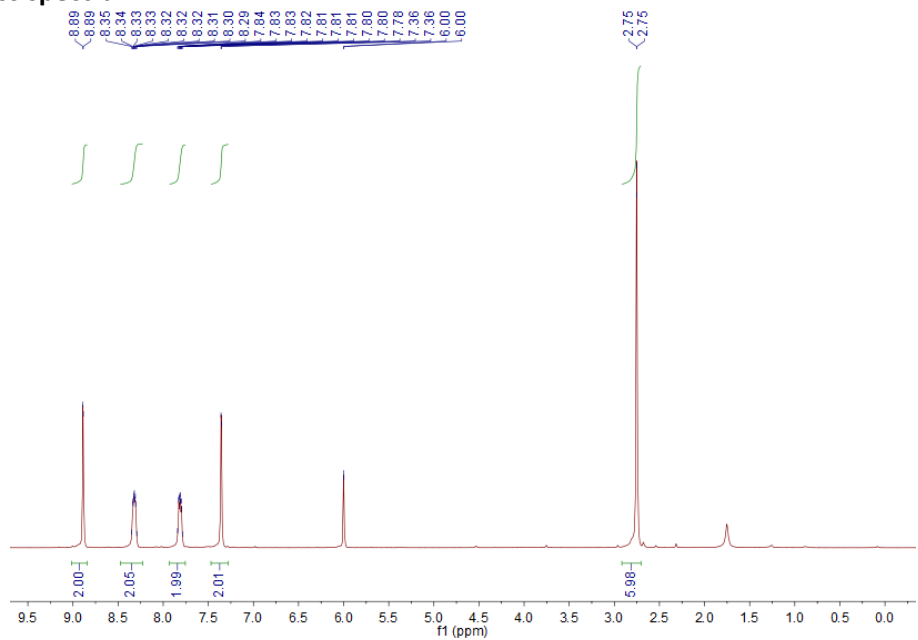

**Figure S25**
$^1$H NMR spectrum of **5** (300 MHz, $C_2D_2Cl_2$, 298 K).

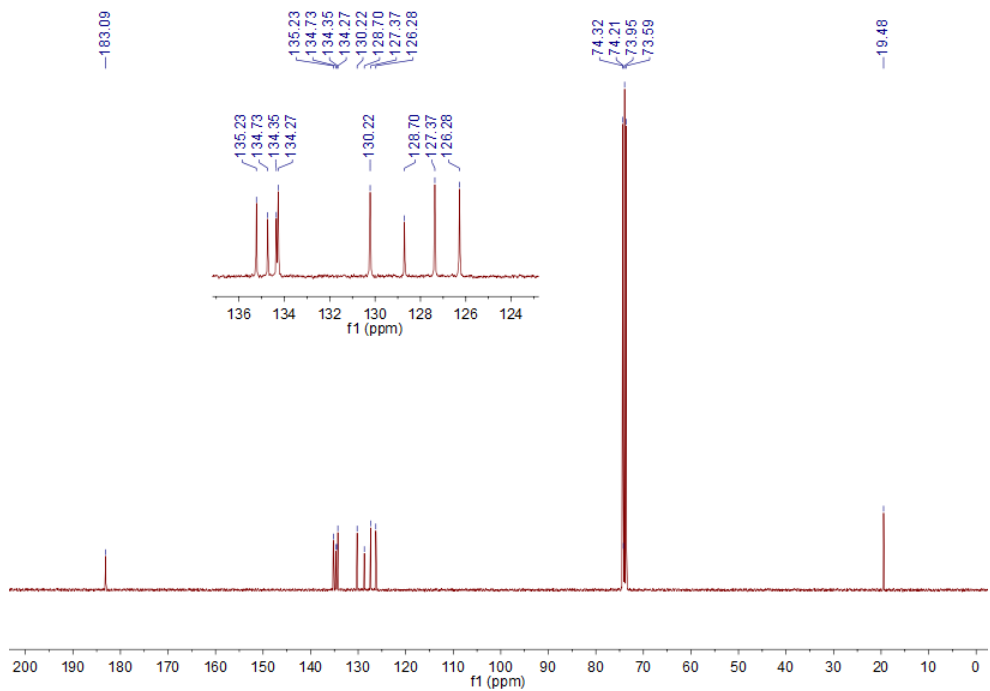

**Figure S26**
$^{13}$C NMR spectrum of **5** (75 MHz, $C_2D_2Cl_4$, 298 K).

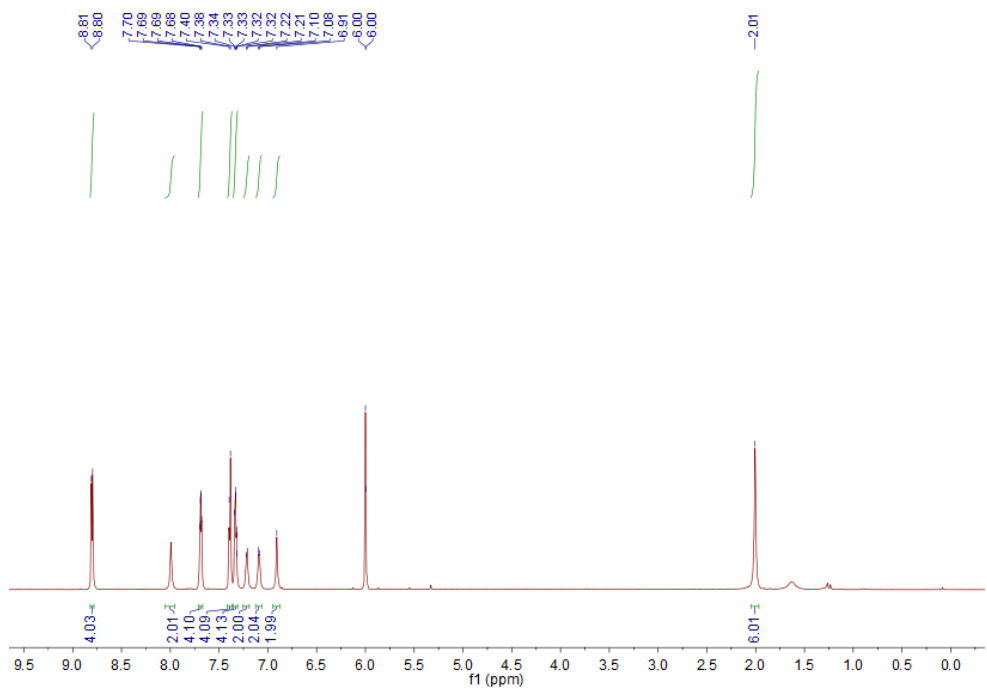

**Figure S27**
$^{1}$H NMR spectrum of monomer **1** (300 MHz, $C_2D_2Cl_2$, 298 K).

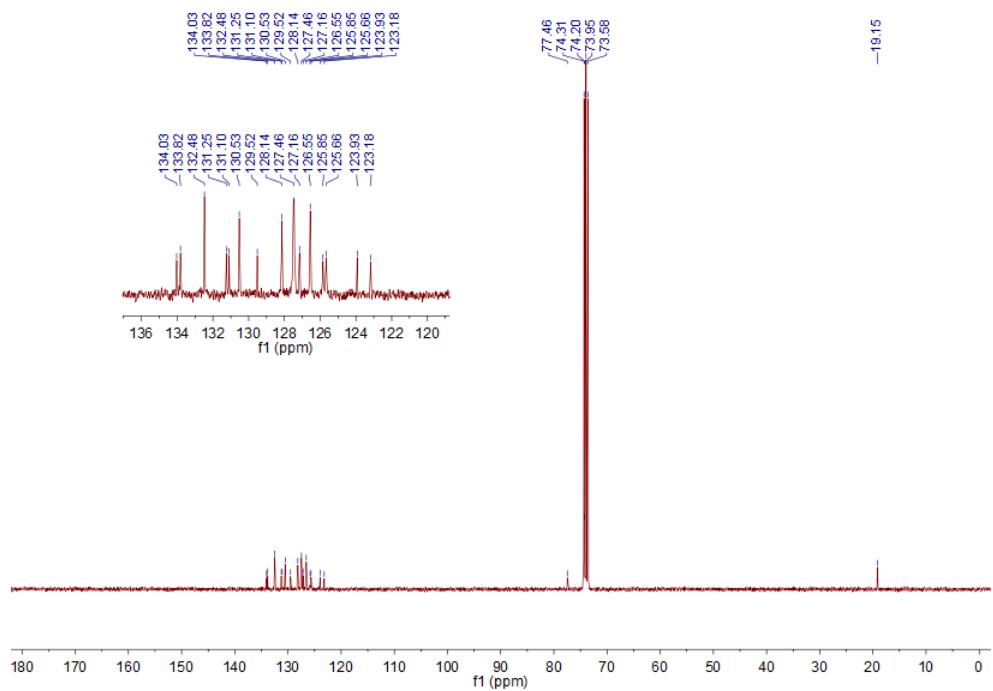

**Figure S28**
$^{13}$C NMR spectrum of monomer **1** (75 MHz, $C_2D_2Cl_4$, 298 K).

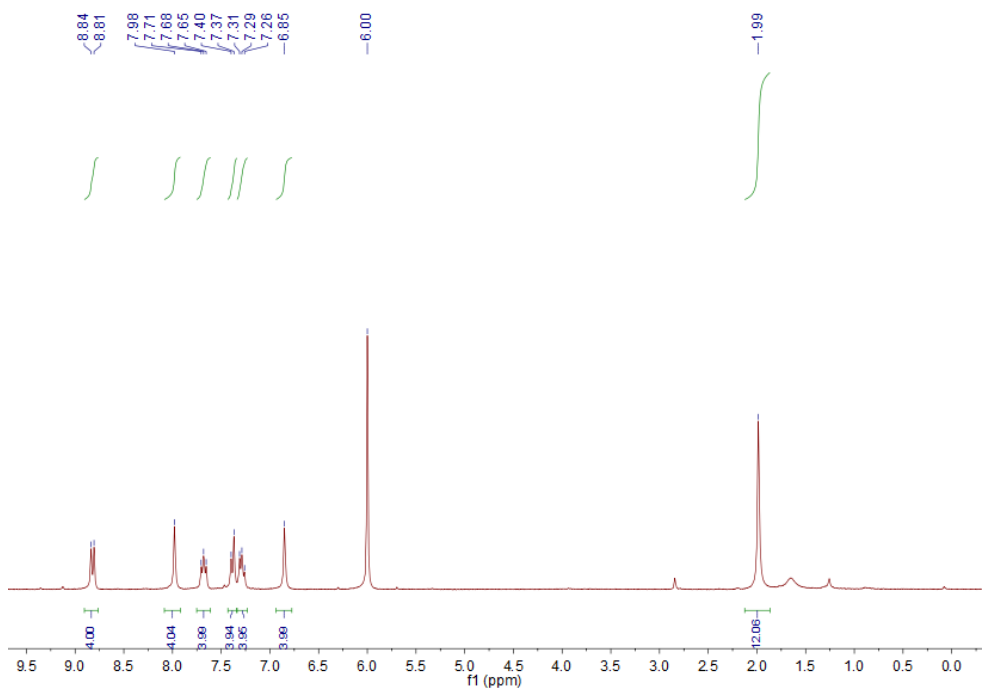

**Figure S29**
¹H NMR spectrum of monomer **3** (300 MHz, C$_2$D$_2$Cl$_2$, 298 K).

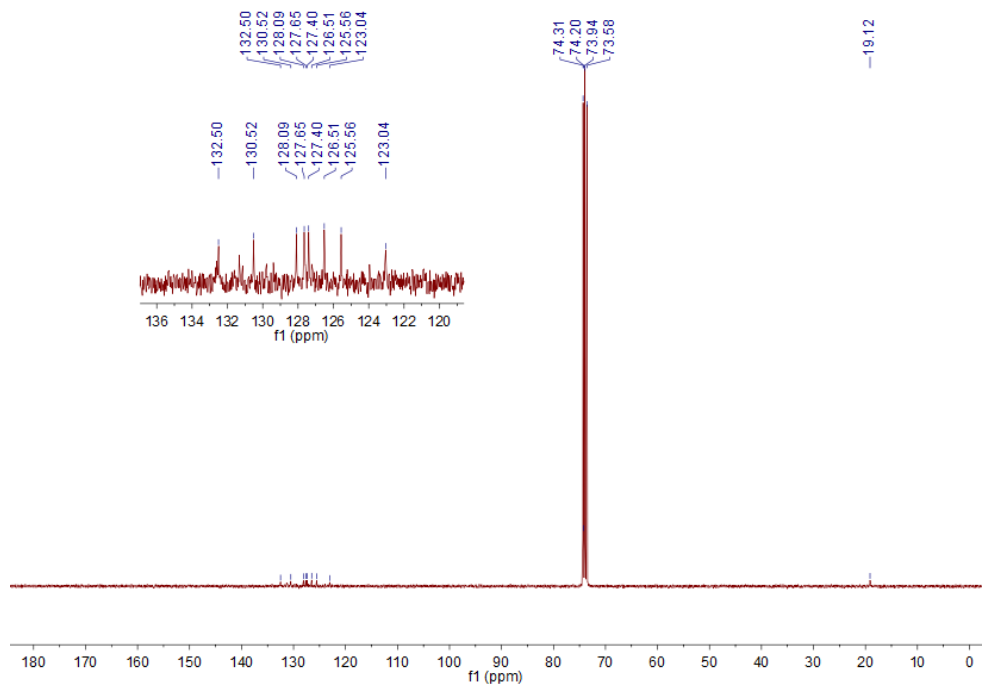

**Figure S30**
¹³C NMR spectrum of monomer **3** (75 MHz, C$_2$D$_2$Cl$_4$, 298 K).

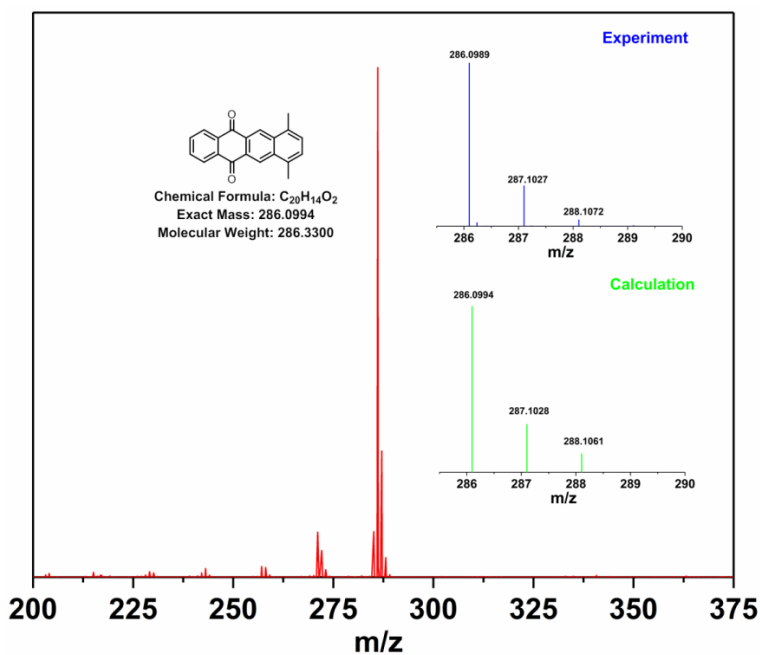

**Figure S31**
High-resolution MALDI-TOF mass spectrum of **7**. Inset displays the isotopic distribution in comparison to the simulated pattern.

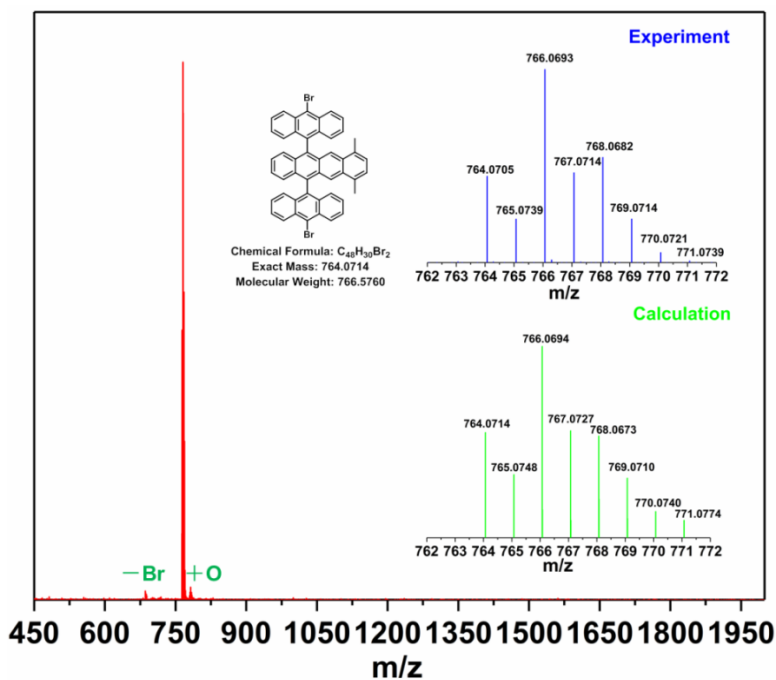

**Figure S32**
High-resolution MALDI-TOF mass spectrum of monomer **1**. Two small signals from [M–Br]$^+$ and [M+O]$^+$ were also detected most probably due to debromination and oxidation, respectively, under the measurement condition or during sample preparation. Inset displays the isotopic distribution in comparison to the simulated pattern.

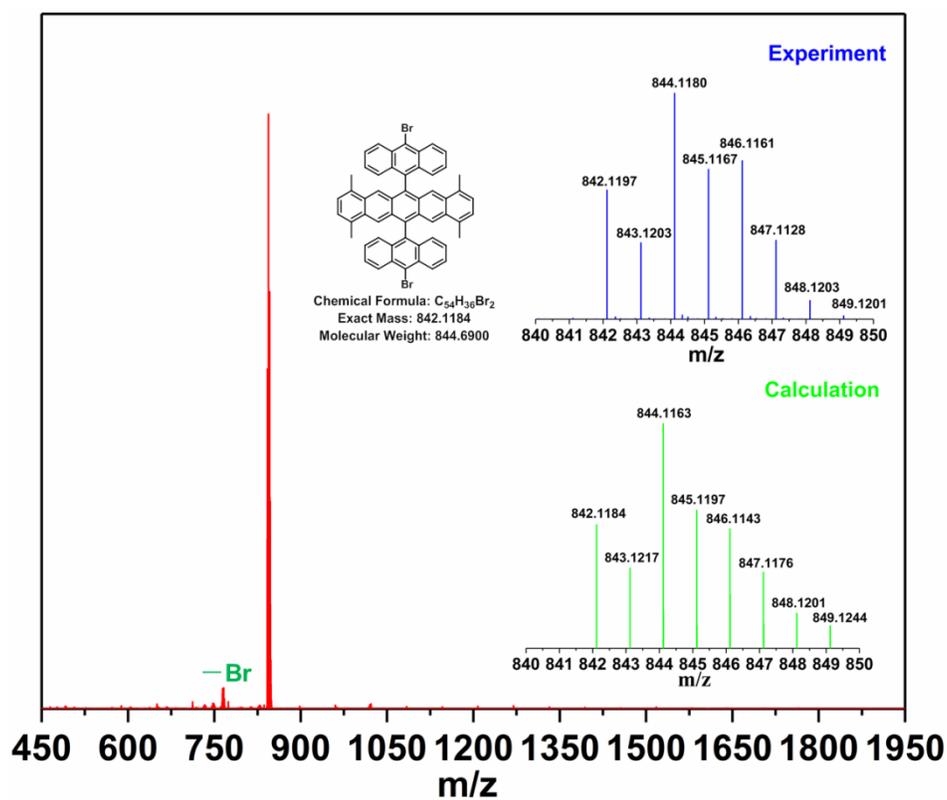

**Figure S33**

High-resolution MALDI-TOF mass spectrum of monomer **3**. A small signal from [M–Br]$^+$ was also detected most probably due to debromination under the measurement condition. Inset displays the isotopic distribution in comparison to the simulated pattern.